\documentclass[aps,prd,reprint,floatfix]{revtex4-1}
\usepackage{placeins}
\usepackage[utf8]{inputenc}
\usepackage[compatibility=false]{caption}
\usepackage{placeins}
\usepackage{blindtext}
\usepackage{amsmath}
\usepackage{graphicx}
\usepackage{stmaryrd}
\usepackage{tcolorbox}
\usepackage{amsfonts}
\usepackage{appendix}
\usepackage{subcaption}
\usepackage{natbib}
\usepackage{array}
\usepackage{multirow}
\usepackage{xcolor}
\usepackage{float}
\usepackage{pifont}
\newcommand{\cmark}{\ding{51}}
\newcommand{\xmark}{\ding{56}}

\usepackage{hyperref}

\begin{document}

\title{Two times or none?}
\author{Michael Ridley}
\email{mikeridleyphysics@gmail.com}
\affiliation{Faculty of Engineering and the Institute of Nanotechnology and Advanced Materials,
Bar-Ilan University, Ramat Gan 5290002, Israel}

\author{Eliahu Cohen}
\affiliation{Faculty of Engineering and the Institute of Nanotechnology and Advanced Materials,
Bar-Ilan University, Ramat Gan 5290002, Israel}

\begin{abstract}
Attempts to treat time on an equivalent footing with space in quantum mechanics have been apparently dominated by `timeless' approaches, such as the one of Page and Wootters, which allow meaningful discussion of a `time operator'. However, there is an alternative, and significantly less studied approach, due to Bauer, which makes use of the `pseudospin' extension of the state space, effectively adding a backwards-time degree of freedom. This two-time approach allows definition of a `time operator' and moreover  bears interesting relations with other time-symmetric formulations of quantum mechanics. We review and compare these approaches to quantum time, emphasizing that there is a subtle choice between the timeless framework and the two-time approach. Finally, we sketch a framework in which the timeless philosophy can be combined with two-time quantum mechanics. 
\end{abstract}

\maketitle

\section{Introduction}

In the standard approach to quantum mechanics, time is not a dynamical variable. Rather, it is treated as a Newtonian `background' parameter of the theory. In contrast with the dynamical treatment of spatial coordinates, no Hermitian operator is assigned to time. Indeed, introducing an ideal time operator canonically conjugate to the Hamiltonian would render the latter unbounded from below. This objection to the existence of a time operator is attributed to Pauli \cite{pauli1933allgemeinen}, and was later strengthened by Unruh and Wald \cite{unruh_time_1989}.

This way of viewing time as a background parameter stands in stark contrast with the picture of relativistic spacetime theories, in which time is treated as one of the dimensions in a unified spacetime geometry \cite{horwitz_two_1988}. The geometric concept of time includes all moments, or spacetime events, in all relativistic reference frames, within a 4-dimensional block-universe picture \cite{reichenbach2012philosophy,vaccaro2018quantum,silberstein2018beyond}. The block universe view is usually married with an \emph{eternalist} theory of time, in which past, present and future are taken to be equally real \cite{isham_continuous_1995}.

This tension between the two pictures of time has been allowed to remain in place, at the foundations of the two most successful theories in physics -- general relativity and quantum mechanics. This is the so-called `problem of time' \cite{anderson2012problem,altaie2022time}. Recent work suggests that, rather than building a general-relativistic theory of quantum mechanics in the usual language of quantum systems, efforts should be refocused on a quantum theory which retains the event-based ontology of general relativity \cite{gambini2018montevideo,maccone_fundamental_2019}. In one theory resulting from this program -- geometric event-based (GEB) quantum mechanics -- spacetime operators are introduced which are conjugate to the 4-momentum, thereby describing space and time with formal similarity \cite{giovannetti2023geometric}. Another approach is provided by Vaccaro, who shows that time-symmetry violation leads to the picture of states as localized in space but spread out over all times, and therefore that the asymmetry between spatial and temporal coordinates underlying the problem of time might not be fundamental \cite{vaccaro2016quantum}.  

In the non-relativistic case, time operators were introduced which make use of an external `clock' system \cite{maccone2020quantum}, an idea which goes back to Page and Wooters (henceforth the PW approach) \cite{page1983evolution}. This is the so-called `timeless' framework, as it starts from consideration of a closed universe, illustrated schematically in Fig. \ref{fig:PW_schematic}, whose wavefunction cannot evolve in time in the usual way due to a Hamiltonian constraint. As such, it is currently an actively studied pathway \cite{marletto2017evolution,foti2021time,baumann2022noncausal,adlam2022watching,rijavec2023robustness,stoica2024empirical,kuypers2024measuring} for introducing a quantum theory of time. Some have raised questions regarding the validity of the decomposition into subsystems that is used by the PW approach \cite{hohn2021equivalence}, as well as the correct interpretation of quantum mechanics which accompanies it \cite{adlam2022watching}. Moreover, Kucha\v{r} in Ref. \cite{kuchavr2011time} made a number of criticisms of the PW approach to quantum gravity. These centered around the probabilities derived from the PW formalism, which appear to violate the constraint equations, and to give inconsistent results for multiple time measurements. Efforts to overcome these criticisms have invoked the `evolving constants’ approach \cite{rovelli_time_1991} parametrized such that correct propagators arise \cite{gambini2009conditional} or carefully formalized von Neumann measurements \cite{giovannetti2015quantum}, and are still ongoing \cite{hohn2021trinity,hausmann2025measurement}.

An alternative, but much less-studied approach due to Bauer \cite{bauer1983time,bauer2014dynamical,bauer2017problem}, suggests to enlarge the state space to include both forwards- and backwards-time processes, leading to a Hermitian time operator which is conjugate to the Hamiltonian. The inclusion of a past-oriented temporal degree of freedom either (i) explicitly, in the mathematical structure of the theory, or (ii) implicitly, in the interpretation of the theory, has a long history in quantum mechanics. There have been several explicitly retrocausal formulations of quantum mechanics which carry out (i), such as the two-state vector formalism (TSVF) \cite{aharonov_time_1964,aharonov_two-state_2008} and its multiple-time extensions \cite{aharonov_multiple-time_2009,aharonov_each_2014}, or (ii), which include the transactional interpretation (TI) \cite{cramer1980generalized,cramer_transactional_1986,cramer_overview_1988} and its variants \cite{kastner2012possibilist,kastner_transactional_2013,kastner2016transactional,kastner2017there}. Other explicitly retrocausal quantum theories can be found in Refs. \cite{watanabe_symmetry_1955,davies1970quantum,davies1971extension,davies1972extension,davidon1976квантовая,sutherland1983bell,reznik1995time,wharton_time-symmetric_2007,pegg2008retrocausality,price2008toy,sutherland2008causally,argaman_bells_2010,heaney2013symmetrical,evans2015retrocausality,price2015disentangling,castagnoli2016relation,sutherland2017lagrangian,wharton2018new,bopp2019bi,sen2019local,sen2020effect,cohen2020realism1,cohen2020realism2,bopp2020avoid,drummond2021time,ridley2023quantum,wharton2024localized,telali2025causality,Ho2025}. In addition, recent work suggests that, even at the macroscopic level, two emergent opposing arrows of time can be derived from the underlying quantum dynamics (modelled as a Markovian open quantum system) \cite{guff2025emergence}. This is closely related to another work which models superpositions of thermodynamic arrows of time \cite{rubino2021quantum} or distinct causal orderings \cite{oreshkov_quantum_2012,chiribella2022quantum,cao2022quantum,mothe2024reassessing} using the process matrix formalism. Such superpositions were experimentally realized and studied in Refs. \cite{guo2024experimental,rozema2024experimental,stromberg2024experimental}. Some of these approaches, like the TI inherently rely on emission/absorption corresponding to retarded/advanced solutions of relativistic equations of motion, while others like the TSVF perfectly align with non-relativistic quantum mechanics. Finally, we note that within many-body quantum field theories, another time symmetric approach has been developed, which makes use of a doubled time axis, known as the Keldysh contour \cite{keldysh_diagram_1964,stefanucci2025nonequilibrium}, for the propagation of quantum statistical mechanical expectation values out of equilibrium. This extended time contour is shown in Fig. \ref{fig:Time_contour}.

Most of these theories can be understood in the realist sense, as a representational model of real processes occurring in nature, but differ insofar as they take the retrocausal part of those processes to be physical. It was recently argued at length that representational models exhibiting retrocausality should also exhibit the property of \emph{event symmetry} -- the basic description of the system at each time point is structurally identical; there should be no ontologically privileged points in time \cite{ridley2025time}. Moreover, the type of retrocausality exhibited by an event-symmetric model should be `all-at-once' \cite{adlam2022two} -- it should treat all times in a history sequence simultaneously, rather than modeling the system with dynamical waves propagating from past/future boundary conditions. Such theories exhibit a type of retrocausation which is not mediated by a record of future events that exists at the measurement time; rather, a form of `mutual causation' obtains, where the past and future are influencing each other at all times in an interchangeable manner \cite{adlam2018spooky,stoica2021post}.

Temporal structure and probability assignments can be intimately related, connecting ontology with measurement \cite{hellmann2007multiple,ridley2023quantum, ridley2025MRW}. In a global, block universe approach to time, time-localized quantum states must be replaced with histories \cite{griffiths_consistent_1984}, and probability assignments to entire histories in the global wavefunction must be made \cite{ridley2025time}. It is notable in this connection that one derivation of the Born measure -- Zurek's `envariance'-based approach -- makes use of a decomposition of the universal wavefunction into subsystems that is similar to the decomposition taken in the PW approach \cite{zurek_environment-assisted_2003,zurek_probabilities_2005,vaidman_derivations_2020}. Whereas the former makes use of entanglement to ensure invariance of a probability measure under symmetry transformations of the composite system+environment state, the latter uses entanglement between subsystems to define the passage of time itself \cite{wootters1984time,moreva2014time}. This `coarse-graining' step is inherently problematic, for both methods. It would be desirable to construct an event-symmetric theory that avoids this kind of decomposition, while simultaneously yielding both a time operator and the Born rule.

This paper proceeds as follows. Section \ref{sec:PW} contains an overview of the PW approach. Bauer's approach to quantum mechanics and the problem of time is outlined in Section \ref{sec:Bauer}, where we also introduce the concept of the `present' as `pinched' in between corresponding time values on the two branches of the Keldysh contour. Next, we summarize four comparable time-extended formulations of quantum mechanics in Section \ref{sec:TimeQM}, namely the decoherent histories approach, the TSVF and its multiple-time generalization, the transactional interpretation, and the fixed-point formulation (FPF). In Section \ref{sec:Comparison}, we analyze the mathematical and conceptual connections between the PW approach and the various retrocausal approaches considered previously. Motivated primarily by the intention to make quantum mechanics conceptually compatible with general relativity, we are thus able to create a taxonomy of temporal properties satisfied by the different quantum formulations, thereby offering a comparison of two-time quantum mechanics and the timeless formulation. Finally, we propose a way to \emph{combine} the two-time and timeless approaches considered in this work. 

\section{Page-Wootters Approach}\label{sec:PW}

\begin{figure}
    \centering
    \includegraphics[width=\linewidth]{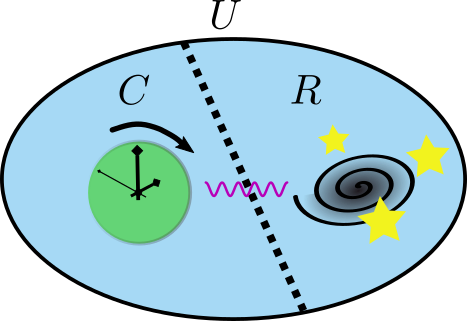}
    \caption{A schematic representation of the division of the universe $U$ into the `clock' subsystem $C$ and the `rest of the universe' $R$, within the PW approach. The wiggly purple line indicates the presence of an interaction between $C$ and $R$.}
    \label{fig:PW_schematic}
\end{figure}

The timeless framework of quantum mechanics, as formulated by Page and Wootters \cite{page1983evolution} and later broadened in \cite{giovannetti2015quantum} introduces a clock subsystem, whose state is given by a vector in a Hilbert space $\mathcal{H}_C$, and the rest of the system, represented by a state in a Hilbert space $\mathcal{H}_R$, whose evolution is studied from the perspective of the clock subsystem. The joint system $|\Psi\rangle\rangle$ exists in the \emph{kinematical} Hilbert space \begin{equation}
    \mathcal{H}_{kin}\equiv\mathcal{H}_C\otimes\mathcal{H}_R.
\end{equation}

It is assumed to be a closed system and hence it is governed by the following equation
\begin{equation}
    H_T|\Psi\rangle\rangle = 0,
    \label{eq:constraint}
\end{equation}
where $H_T$ is the total Hamiltonian acting in $\mathcal{H}_{kin}$. For this reason, we may describe $|\Psi\rangle\rangle$ as a \emph{clock neutral} object - it describes the physical situation prior to choosing a temporal reference frame \cite{hohn2020switch}. The division of the `Universe' into these subsystems is illustrated schematically in Fig. \ref{fig:PW_schematic}. 

Eq. \eqref{eq:constraint} is mathematically analogous to the Wheeler-DeWitt (WDW) equation which, in fact, has wider applicability as a Hamiltonian constraint in canonical quantum gravity \cite{dewitt1967quantum,hartle1983wave,isham1993canonical,Rovelli_2015}. This equation also implies that the total system $|\Psi\rangle\rangle$, originally being thought of as the `Universe' \cite{page1983evolution}, does not evolve with respect to an external time -- this is the so-called `frozen formalism problem' (which will be shown to be lifted once we condition on quantum clock outcomes). Note that the imposition of $|\Psi\rangle\rangle$ being an eigenstate with a null eigenvalue by Eq. \eqref{eq:constraint} is not as restrictive as it may appear to be \cite{giovannetti2015quantum} since Hamiltonians which differ by constant terms are physically equivalent.

The clock system typically has a time operator $T_C$ associated with its time. It is desirable that $H_C$ generates translations in time (a property often called covariance in the quantum time literature), i.e. 

\begin{equation}
    |t_0+t_C\rangle = e^{-i H_C t_C/\hbar} |t_0\rangle,
\end{equation}
where $|t_0\rangle$ and $|t_0+t_C\rangle$ are taken to be clock states, but this does not necessarily require $T_C$ to be self-adjoint and canonically conjugate to $H_C$. Indeed, it was shown how to construct a quantum clock as a covariant positive operator-valued measure (POVM), which means that the clock states are not necessarily eigenstates of $T_C$ \cite{busch1995operational, busch2016quantum, loveridge2019relative,hohn2021trinity}. The first-moment operator of the POVM is symmetric but, for a semibounded spectrum, cannot be self-adjoint. While such clocks cannot mark time as sharply as \textit{ideal clocks}, obeying $[T_C,H_C]=i\hbar I$, $T_C|t_C\rangle = t_C|t_C\rangle$, and $H_C=-i\hbar \partial/\partial t_C$, they directly bypass Pauli's objection which only rules out self-adjoint canonical time operators. Nevertheless, we assume here for simplicity the case of ideal clocks. Although achieving an operationally accessible unbounded spectrum is unrealistic for practical clocks, ideal clocks often help us avoid technicalities associated with real clocks \cite{salecker1958quantum, peres1980measurement}, while providing useful approximations thereof \cite{hartle1988quantum, singh2018modeling}. In any case, the PW approach avoids Pauli’s obstruction either by (i) choosing an ideal clock with an unbounded Hamiltonian, so a conjugate operator exists, or (ii) using a POVM clock, where self-adjointness is not required \cite{cafasso2024quantum}.

Still, the analysis of ideal clocks could be a bit elaborate and intricate due to interactions between the subsystems \cite{paiva2022flow,paiva2022non,rijavec2023robustness}, illustrated schematically by the red wiggly line in Fig. \ref{fig:PW_schematic}. Let $H_R$ denote again the Hamiltonian of the system of interest and let $H_{int}(T_C)$ represent the time-dependent term of the evolution of system $R$ set by clock $C$, which is an interaction between $C$ and $R$. In this case we have
\begin{equation}
    H_T = H_C \otimes \mathbf{I}_{R} + \mathbf{I}_{C} \otimes H_R + H_{int}(T_C),
    \label{eq:pw-h}
\end{equation}
where $H_{int}$ is assumed to be independent of $H_C$. $\mathbf{I}_{R}$ and $\mathbf{I}_{C}$ denote identity operators in the corresponding subspaces. Inserting Eq. \eqref{eq:pw-h} into Eq. \eqref{eq:constraint}, applying a scalar product by an eigenstate $|t_C\rangle$ of $T_C$ on the left, and defining the time-dependent wavevector of $R$ as 

\begin{equation}
|\psi_{R}(t_C)\rangle \equiv \left(\langle t_C| \otimes \mathbf{I}_{R} \right)|\Psi\rangle\rangle,
\end{equation}

we obtain 

\begin{equation}
    i\hbar \frac{\partial}{\partial t_C} |\psi_{R}(t_C)\rangle = H_R |\psi_{R}(t_C)\rangle + \int dt' K\left(t_C,t'\right)|\psi_{R}(t')\rangle,
    \label{eq:schrod-K-general}
\end{equation}

where $K\left(t_C,t'\right)\equiv\langle t_C | H_{int} | t' \rangle$ is a time-nonlocal term resulting from the interaction. Mathematically, the role of $K\left(t_C,t'\right)$ is to \emph{complete} the projection onto the $R$ subspace, as only the part of $H_{int}$ which acts in the clock subspace acts on the time states in the inner product defining $K\left(t_C,t'\right)$. In general, the solution to Eq. \eqref{eq:schrod-K-general} requires knowledge of the wavefunction $|\psi_{R}(t_C)\rangle$ at all times. In practice, progress can be made in solving Eq. \eqref{eq:schrod-K-general} as a perturbative series in the interaction strength \cite{smith2019quantizing}. Alternatively, the kernel function $K\left(t_C,t'\right)=K\left(t',t_C\right)^{\dagger}$ is proportional to a delta function $\delta\left(t_C-t'\right)$ in the case where $H_{int}$ is time-diagonal (assuming the completeness relation $\langle t_{1}|t_{2} \rangle = \delta\left(t_{1}-t_{2}\right)$ is valid). This results in a time-local form for the equation of motion

\begin{equation}
    i\hbar \frac{\partial}{\partial t_C} |\psi_{R}(t_C)\rangle = \left[H_R + H_{int}(t_C)\right]|\psi_{R}(t_C)\rangle,
    \label{eq:schrod-general}
\end{equation}

which is the time-dependent Schr\"odinger equation denoting the evolution of system $R$ with respect to the time measured by clock $C$. Then, the usual unitary evolution of a quantum system can be recovered from the static, timeless picture. As a result, $|\Psi\rangle\rangle$ can be expressed as
\begin{equation}\label{eq:PW_History}
    |\Psi\rangle\rangle = \int dt_C \, |t_C\rangle \otimes |\psi_{R}(t_C)\rangle,
\end{equation}

which is an example of a Schmidt decomposition in the combined $C+R$ Hilbert space \cite{cafasso2024quantum}.

We note that this procedure, of obtaining an evolution for the `reduced' wavefunction of a subsystem $R$ is mathematically and conceptually similar to that carried out as a matter of routine for open quantum systems such as those considered in quantum electronics \cite{stefanucci2025nonequilibrium}. The main difference between such formalisms and the PW approach is that the former carry out an \emph{embedding} of the (usually smaller) quantum system within the global structure, leading to reduced equations of motion for this system. The PW approach, by contrast, carries out the inverse procedure known as \emph{inbedding} via conditioning on the clock to yield reduced equations of motion for the rest of the global structure. Both embedding and inbedding lead to modified, and sometimes non-unitary effective dynamics for the subsystem of interest. This suggests that sophisticated many-body open quantum system techniques such as the non-equilibrium Green's function (NEGF) \cite{stefanucci2025nonequilibrium} method can be applied to PW models, although to our knowledge this connection has not yet been fully developed in the literature. 

Since $|\Psi\rangle\rangle$ contains information about $|\psi_{R}(t_C)\rangle$ at every $t_C$, it is sometimes referred to as the \textit{history state} \cite{diaz2019history}. As such, we can view the universal wavefunction $|\Psi\rangle\rangle$ as `timeless' from the birds-eye, atemporal perspective, or as `time-full' from the perspective of the spatially-localized clock, existing at \emph{all} times. Operationally, one first computes the joint statistics of the chosen system observable with the clock reading, and for sequences of events simply interleaves these clock effects with the usual time-ordered system evolutions; this recipe gives the standard Born probabilities when the clock is ideal and remains accurate—up to corrections set by the finite time resolution—when the clock is described by a covariant POVM \cite{hohn2021trinity}.

In the relativistic Page–Wootters setting, an apparent loophole in Pauli’s theorem arises because Dirac- and Klein–Gordon-type Hamiltonians possess a two-sided, unbounded energy spectrum, making a self-adjoint time operator formally admissible and allowing an ideal clock whose positive- and negative-frequency branches generate forward and backward relational dynamics. However, the ensuing act of mixing positive and negative frequencies within the relativistic approach may conflict with charge superselection unless we restrict ourselves only to positive frequencies and then any clock that remains within the physical sector again obtains a lower bound on energy and Pauli’s obstruction re-emerges. Operationally, one therefore retreats from a sharp time-energy canonical pair to a covariant time POVM whose first-moment operator is merely symmetric, not self-adjoint as discussed above in the non-relativistic case. Timing accuracy is again traded for finite energy spread. The relativistic framework thus clarifies, rather than abolishes, Pauli’s lesson: unbounded spectra make an ideal clock conceivable, but once physical constraints such as charge superselection, stability and field quantization are enforced, the clock should better be non-ideal, described by a coherent POVM \cite{hohn2021equivalence} with controllable but non-vanishing indeterminacy, exactly mirroring the resource-limited clocks of the non-relativistic PW theory.

\section{Bauer's Approach}\label{sec:Bauer}
In this section, we follow the derivation of Bauer in Ref. \cite{bauer1983time} of a time operator satisfying the appropriate commutation relations. This derivation involves an extension of the usual energy state space in two equivalent ways; (i) the negative energy extension and (ii) the pseudospin extension which introduces time orientation as a degree of freedom.
Take a system described by Hamiltonian $H$:

\begin{equation}\label{eq:Hamiltonian}
    H\left|E\right\rangle =E\left|E\right\rangle ,\ E\geq0
\end{equation}

The eigenstates satisfy the inner product:

\begin{equation}
    \left\langle E\right|\left.E'\right\rangle =\delta\left(E-E'\right)
\end{equation}

We now define the \emph{energy shift operator}

\begin{equation}\label{eq:D}
    D\left(\varepsilon\right)\left|E\right\rangle =\theta\left(E-\varepsilon\right)\left|E-\varepsilon\right\rangle, 
\end{equation}

and its adjoint

\begin{equation}\label{eq:D^+}
    D^{\dagger}\left(\varepsilon\right)\left|E\right\rangle =\theta\left(E+\varepsilon\right)\left|E+\varepsilon\right\rangle =\left|E+\varepsilon\right\rangle 
\end{equation}

These operators (understood as acting on the ket $\left|E\right\rangle $) satisfy

\begin{equation}
    D\left(\varepsilon\right)D^{\dagger}\left(\varepsilon\right)	=\mathbf{I} ,\  
    D^{\dagger}\left(\varepsilon\right)D\left(\varepsilon\right)	=\theta\left(E-\varepsilon\right),
\end{equation}

so $D\left(\varepsilon\right)$ is \emph{not} unitary.

From the lower-bounded Hamiltonian defined in equation \eqref{eq:Hamiltonian}, it is possible to construct the extended state space in a spinor state construction:

\begin{equation}
    \left|\tilde{E}\right\rangle =\begin{cases}
\begin{array}{c}
\left[\begin{array}{c}
\left|E\right\rangle \\
0
\end{array}\right],\ E\geq0\\
\left[\begin{array}{c}
0\\
\left|-E\right\rangle 
\end{array}\right],\ E\le 0
\end{array}\end{cases}
\end{equation}

Introducing the operator 

\begin{equation}
    Q=\left(\begin{array}{cc}
1 & 0\\
0 & -1
\end{array}\right),
\end{equation}

we construct the extended-space Hamiltonian

\begin{equation}
    \tilde{H}=HQ=\left(\begin{array}{cc}
H & 0\\
0 & -H
\end{array}\right),
\end{equation}

which satisfies

\begin{equation}
    \tilde{H}\left|\tilde{E}\right\rangle =E\left|\tilde{E}\right\rangle ,\ -\infty<E<\infty.
\end{equation}

For completeness, we note that the spinor states $\left|\tilde{E}\right\rangle$ exist in the Hilbert space

\begin{equation}\label{eq:Hilbert_Bauer}
    \mathcal{H}_{B} = Sp\{|E\rangle\}_{E\ge0}\bigoplus Sp\{|E\rangle\}_{E\le 0},
\end{equation}
where `$Sp$' denotes the span of the eigenspectrum defined by Eq. \eqref{eq:Hamiltonian}. We may refer to $\mathcal{H}_{B}$ as the \emph{Bauer space}. We now define the energy shift operator in this extended space as 

\begin{equation}\label{eq:t_DP}
    \tilde{D}\left(\varepsilon\right)=\left(\begin{array}{cc}
D\left(\varepsilon\right) & 0\\
P\left(\varepsilon\right) & D^{\dagger}\left(\varepsilon\right)
\end{array}\right),
\end{equation}

where 

\begin{equation}
    P\left(\varepsilon\right)\equiv\overset{\infty}{\underset{-\infty}{\int}}dE\left|\varepsilon-E\right\rangle \theta\left(\varepsilon-E\right)\theta\left(E\right)\left\langle E\right|=P^{\dagger}\left(\varepsilon\right)
\end{equation}
is an Hermitian operator. One can show that the operator $D\left(\varepsilon\right)$ satisfies the following properties:

\begin{eqnarray}
    \tilde{D}\left(\varepsilon\right)\left|E\right\rangle 	=\left|E-\varepsilon\right\rangle ,\ -\infty<E<\infty \\
    \tilde{D}^{\dagger}\left(\varepsilon\right)\left|E\right\rangle 	=\left|E+\varepsilon\right\rangle ,\ -\infty<E<\infty \\
    \tilde{D}\left(\varepsilon\right)\tilde{D}^{\dagger}\left(\varepsilon\right)	=\tilde{D}^{\dagger}\left(\varepsilon\right)\tilde{D}\left(\varepsilon\right)=\mathbf{I} \\
    \tilde{D}^{\dagger}\left(\varepsilon\right)	=\tilde{D}\left(-\varepsilon\right)=\tilde{D}^{-1}\left(\varepsilon\right) \\
\tilde{D}\left(0\right)=\mathbf{I} \\
    \tilde{D}\left(\varepsilon_{1}\right)\tilde{D}\left(\varepsilon_{2}\right)	=\tilde{D}\left(\varepsilon_{1}+\varepsilon_{2}\right)
\end{eqnarray}

Since the set $\left\{ \tilde{D}\left(\varepsilon\right):\varepsilon\in\mathbb{R}\right\} $ forms a strongly continuous one-parameter unitary group, Stone's theorem guarantees the existence of an Hermitian operator $\hat{t}$ such that 

\begin{equation}\label{eq:Stone}
\tilde{D}\left(\varepsilon\right)=e^{-i\varepsilon\hat{t}}
\end{equation}

The operator $\hat{t}$ is said to be the infinitesimal generator of the group of energy shifts \cite{stone1932one}. From Eq. \eqref{eq:Stone}, the following commutation relation is derivable:

\begin{equation}\label{eq:commutator_1}  
    \left[\tilde{D}\left(\varepsilon\right),\tilde{H}\right]=\varepsilon\tilde{D}\left(\varepsilon\right)
\end{equation}

Expanding both sides of Eq. \eqref{eq:commutator_1} in powers of $\varepsilon$, it is then straightforward to obtain the crucial commutator

\begin{equation}\label{eq:commutator_2}
    \left[\hat{t},\tilde{H}\right]=i
\end{equation},

from which it follows that $\hat{t}$ has a continuous spectrum with delta-normalized eigenvectors:

\begin{equation}\label{eq:t_eigenbasis}
        \hat{t}\left|\tilde{t}\right\rangle 	=t\left|\tilde{t}\right\rangle ,\ -\infty<t<\infty
\end{equation}
\begin{equation}\label{eq:t_delta}
        \left\langle \tilde{t}'\right|\left.\tilde{t}\right\rangle 	=\delta\left(t'-t\right)
\end{equation}

Thus, the time operator in Eq. \eqref{eq:t_operator} can be expanded in the time eigenbasis defined by Eqs. \eqref{eq:t_eigenbasis} and \eqref{eq:t_delta}:

\begin{equation}\label{eq:t_expanded}
    \hat{t}=\overset{\infty}{\underset{-\infty}{\int}}dt\thinspace t\left|t\right\rangle \left\langle t\right|
\end{equation}

By Eqs. \eqref{eq:t_DP} and \eqref{eq:Stone}, another representation of the time operator is given by: 

\begin{equation}\label{eq:t_operator}
    \hat{t}=i\left.\frac{d\tilde{D}\left(\varepsilon\right)}{d\varepsilon}\right|_{\varepsilon=0}=i\frac{d}{d\varepsilon}\left.\left(\begin{array}{cc}
D\left(\varepsilon\right) & 0\\
P\left(\varepsilon\right) & D^{\dagger}\left(\varepsilon\right)
\end{array}\right)\right|_{\varepsilon=0}
\end{equation}

Since the $\varepsilon$-derivative of $P\left(\varepsilon\right)$ vanishes in the limit $\varepsilon \rightarrow 0$ \cite{bauer1983time}, we can also write this in diagonal form:

\begin{equation}\label{eq:t_diagonal}
    \hat{t}=\left(\begin{array}{cc}
\hat{t}^{f} & 0\\
0 & \hat{t}^{b}
\end{array}\right),
\end{equation}

where we define the \emph{forwards-time operator}:

\begin{equation}\label{eq:t_f}
    \hat{t}^{f}\equiv i\left.\frac{dD\left(\varepsilon\right)}{d\varepsilon}\right|_{\varepsilon=0}
\end{equation}

and the corresponding \emph{backwards-time operator}:

\begin{equation}\label{eq:t_b}
    \hat{t}^{b}\equiv i\left.\frac{dD^{\dagger}\left(\varepsilon\right)}{d\varepsilon}\right|_{\varepsilon=0}
\end{equation}

From Eq. \eqref{eq:t_operator}, the inner product relationships can be derived:

\begin{eqnarray}
    \left\langle t\right|\tilde{H}\left|t'\right\rangle =i\frac{\partial}{\partial t}\delta\left(t-t'\right)\\
    \left\langle\tilde{E}\right|\hat{t}\left|\tilde{E'}\right\rangle 	=-i\frac{\partial}{\partial E}\delta\left(E-E'\right)\\
\left\langle\tilde{E}\right|\hat{t}\left|t\right\rangle 	=i\frac{\partial}{\partial E}\left\langle \tilde{E}\right\Vert \left.t\right\rangle =t\left\langle \tilde{E}\right\Vert \left.t\right\rangle ,
\end{eqnarray}

which in turn implies

\begin{equation}
    \left\langle \tilde{E}\right|\left.t\right\rangle =\frac{e^{-iEt}}{\sqrt{2\pi}}
\end{equation}

Now we turn to the `pseudospin extension' of the physical space, with corresponding Hamiltonian

\begin{equation}
    H'=\left(\begin{array}{cc}
H & 0\\
0 & H
\end{array}\right)=\tilde{H}Q.
\end{equation}

Using the commutation relation Eq. \eqref{eq:commutator_2}, we can show that this Hamiltonian satisfies

\begin{equation}\label{commutator_3}
    \left[\tilde{t},H'\right]=iQ
\end{equation}

$H'$ also satisfies the following eigenproblem

\begin{equation}
    H'\left|E\right\rangle =\left|E\right|\left|E\right\rangle ,\ -\infty<E<\infty,
\end{equation}

so it has an eigenspectrum with twofold degeneracy everywhere. Therefore, setting $E=\left|E\right|$, we can write

\begin{equation}\label{eq:pseigenspectrum}
    H'\left|E,\alpha\right\rangle =E\left|E,\alpha\right\rangle ,\ 0\leq E<\infty,
\end{equation}
where $\alpha=f,b$ correspond to the forwards-directed and backwards-directed time labels. To see this more explicitly, we consider the time evolution of any observable $A$ in the usual formulation of quantum mechanics

\begin{equation}
    \frac{d\left\langle A\right\rangle }{dt}=-i\left\langle \left[A,H'\right]\right\rangle +\left\langle \frac{\partial A}{\partial t}\right\rangle ,
\end{equation}

where $\left\langle A\right\rangle \equiv\left\langle \Psi\right|A\left|\Psi\right\rangle $ for any normalized $\left|\Psi\right\rangle $.

Now consider the case where $A=\hat{t}$. Since $\hat{t}$ does not depend explicitly on time, we use Eq. \eqref{commutator_3} to obtain

\begin{equation}\label{eq:EoM}
    \frac{d\left\langle \hat{t}\right\rangle }{dt}=\left\langle Q\right\rangle. 
\end{equation}

Now, given some arbitrary state vector $\left|\Psi\left(t\right)\right\rangle $ in the extended space, we can take the inner product in Eq. \eqref{eq:EoM} with respect to either its $f$ or $b$ components. 

In the former case, this is given by $\left|P\Psi\right\rangle \equiv\left|\Psi^{f}\right\rangle $, where 

\begin{equation}\label{eq:P}
    P=\left(\begin{array}{cc}
1 & 0\\
0 & 0
\end{array}\right).
\end{equation}

Using Eq. \eqref{eq:EoM}, we obtain for this case

\begin{equation}
    d\left\langle \hat{t}\right\rangle ^{f}=dt,
\end{equation}

corresponding to a shift in the forwards-time direction.

In the latter case, the relevant component is given by $\left|\bar{P}\Psi\right\rangle \equiv\left|\Psi^{b}\right\rangle  $, where 

\begin{equation}\label{eq:I-P}
    \bar{P}=\mathbf{I}-P=\left(\begin{array}{cc}
0 & 0\\
0 & 1
\end{array}\right)
\end{equation}

projects onto the $b$ subspace. Substituting this component into Eq. \eqref{eq:EoM}, we obtain

\begin{equation}
    d\left\langle \hat{t}\right\rangle ^{b}=-dt,
\end{equation}

corresponding to a shift in the backwards-time direction.

\begin{figure}[htp]
    \includegraphics[clip, width=\linewidth]{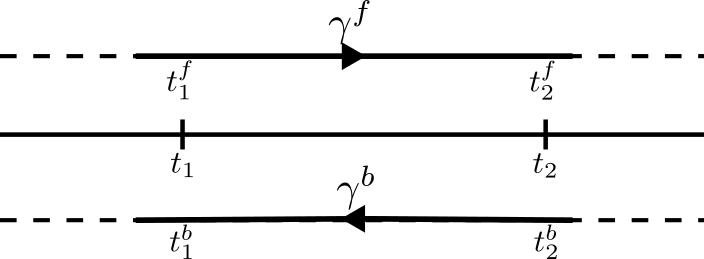}
    \caption{The Keldysh time contour.}\label{fig:Time_contour}
\end{figure}

The pseudo-spin extension can be visualized via a doubling of the time axis to make the Keldysh contour $\gamma$ (shown in Fig. \ref{fig:Time_contour}), which consists of an `upper' branch $\gamma^f$, on which times run from $-\infty$ to $+\infty$, before `turning' to run in the anti-chronological direction from $+\infty$ to $-\infty$ on the `lower' branch $\gamma^b$, such that $\gamma\equiv \gamma^f \oplus \gamma^b$ \cite{keldysh_diagram_1964}. The regular background time $t$ can be thought of as `pinched' exactly at the midpoint of the corresponding branch times $t^f$, $t^b$ in the complex time direction. We choose some very small regularization parameter $\eta$ and set $t^f\left(\eta\right)=t+i\eta$, $t^b\left(\eta\right)=t-i\eta$, so that the measurement time is given by

\begin{equation}\label{eq:pinched_time}
    t=\underset{\eta\rightarrow0}{\lim}\frac{1}{2}\left[t^{f}\left(\eta\right)+t^{b}\left(\eta\right)\right].
\end{equation}

This is suggestive of a time operator, to be discussed later in this work, acting upon states with combined forwards- and backwards-directed parts, to return a single `pinched' present.

Having introduced Bauer’s non-relativistic construction, where a spinor-like `doubling' of a semi-bounded Hamiltonian manufactures a mirror spectrum and thereby revives a self-adjoint time operator, we can now pass to the genuinely relativistic case, in which this mirror already exists in nature. For the free (or minimally coupled) Dirac Hamiltonian, the spectrum spans both positive and negative energies, so Pauli’s lower-bound premise fails automatically and Bauer’s dynamical time operator emerges without any auxiliary degrees of freedom. Mathematically this furnishes an ideal canonical clock, canonically conjugated to the Hamiltonian. Physically, however, the same caveat that limited the doubled non-relativistic model reappears: once charge superselection confines us to the positive-frequency subspace, the time-operator may lose its self-adjointness and reduce to the first moment of a covariant time-POVM, just like in the PW approach, restoring Pauli’s energy-spread versus time-resolution trade-off.

It is important to note that Bauer's approach deals with a continuous energy spectrum, suggesting that there are some difficulties associated with discrete energy scales \cite{bauer1983time}. The PW approach can deal with both, according to the work of Favalli \cite{favalli2020time, favalli2024emergence}, which draws upon earlier results by Pegg \cite{pegg1998complement}.
\section{Time-extended formulations of quantum mechanics}\label{sec:TimeQM}

\subsection{Decoherent Histories}

Throughout the last decades, the development of the field of quantum cosmology has led to the decoherent histories framework \cite{gell-mann_quantum_1996,craig2010consistent,hartle_one_2017}. Unlike the other approaches considered in this Section, this class of formalisms involve no retrocausality. However, it is an important case to consider because it replaces the usual state vector of quantum mechanics with temporally extended states in a Hilbert space constructed from replicas of the usual quantum state space, localized to single points in time. Each time $t$ in some sense involves the instantiation of a `new universe' represented by the Hilbert space $\mathcal{H}_{t}$.

In such theories, the `observer' plays no privileged role, and instead is contained as a subsystem of the global wave function. 

By analogy with the world lines of Minkowski space, macroscopic objects are represented by \emph{histories} - sequences in the space of coarse-grained configurations $\alpha_{1},\alpha_{2},\ldots,\alpha_{N}$, corresponding to a sequence of $N_{t}$ times $t_{1},t_{2},\ldots,t_{N_{t}}$. Here, each $\alpha_{k}$ ranges over a complete basis set of orthogonal outcomes and can be instantiated with a projection operator $P_{\alpha_{k}}\equiv\left|\alpha_{k}\right\rangle \left\langle \alpha_{k}\right|$ satisfying 
\begin{equation}
    \underset{\alpha_{k}}{\sum}P_{\alpha_{k}}=\mathbf{I}
\end{equation}
Given  $\rho_{1}=\left|\psi_{1}\right\rangle \left\langle \psi_{1}\right|$, the initial density matrix of the system at time $t_{1}$, one can then evaluate the probability of a quantum history as:
\begin{gather}
    p\left(\alpha_{N_{t}}\left(t_{N_{t}}\right),\ldots,\alpha_{2}\left(t_{2}\right);\rho_{1}\right)=\\
    \textrm{Tr}\left[P_{\alpha_{N_{t}}}\left(t_{N_{t}}\right)\ldots P_{\alpha_{2}}\left(t_{2}\right)\rho_{1}P_{\alpha_{2}}\left(t_{2}\right)\ldots P_{\alpha_{N_{t}}}\left(t_{N_{t}}\right)\right]
\end{gather}
where $P_{\alpha_{k}}\left(t_{k}\right)\equiv U\left(t_{1},t_{k}\right)P_{\alpha_{k}}U\left(t_{k},t_{1}\right)$ is defined in terms of the unitary time evolution propagator $U\left(t_{k},t_{1}\right)$. In Refs. \cite{wallace_emergent_2012,saunders_many_2010}, this is reformulated in terms of the so-called class operator \cite{riedel_objective_2016}
\begin{equation}
    C_{\mathbf{\alpha}}\equiv\left[P_{\alpha_{N}}\left(t_{N_{t}}\right),\ldots,P_{\alpha_{2}}\left(t_{2}\right)\right]
\end{equation}
which acts on the \textit{history} Hilbert space made up of a product of time-local Hilbert spaces \cite{isham_continuous_1995,isham_continuous_1998}:
\begin{equation}
  \mathcal{H}_{H}\equiv\mathcal{H}_{t_{N_{t}}}\otimes\ldots\otimes\mathcal{H}_{t_{1}}
  \label{eqn:Hilbert_Space_forwards}
\end{equation}  

One can also define `record states', which capture the record of each time snapshot of a single quantum history:

\begin{equation}
    \left|\alpha\right\rangle=C_{\alpha}\left|\psi_{1}\right\rangle
\end{equation}
    
 and the probability of the corresponding quantum history is given by:
    
\begin{gather}
    P\left(\mathbf{\alpha}\right)=\left\langle \psi_{1}\right|C_{\alpha}^{\dagger}C_{\alpha}\left|\psi_{1}\right\rangle = \\
    \left|\left\langle \psi_{1}\right|\left.\alpha_{2}\left(t_{2}\right)\right\rangle \left\langle \alpha_{2}\left(t_{2}\right)\right|\left.\alpha_{3}\left(t_{3}\right)\right\rangle \ldots\left\langle \alpha_{N-1}\left(t_{N-1}\right)\right|\left.\alpha_{N}\left(t_{N}\right)\right\rangle \right|^{2}
\end{gather}

Defining the decoherence functional \cite{wang2025decoherence}
\begin{equation}
    \mathfrak{D}\left(\mathbf{\alpha},\mathbf{\beta}\right)\equiv\left\langle \psi_{1}\right|C_{\alpha}^{\dagger}C_{\beta}\left|\psi_{1}\right\rangle
\end{equation}

we can also write the history probability as 
\begin{equation}
    P\left(\mathbf{\alpha}\right)=\mathfrak{D}\left(\mathbf{\alpha},\mathbf{\alpha}\right)
\end{equation}

and specify the decoherence condition for non-equal decoherent histories $\mathbf{\alpha},\mathbf{\beta}$ as:
\begin{equation}
    \mathfrak{D}\left(\mathbf{\alpha},\mathbf{\beta}\right)=0,\thinspace\mathbf{\alpha}\neq\mathbf{\beta}
\end{equation}

\begin{figure}
  \includegraphics[width=.75\linewidth]{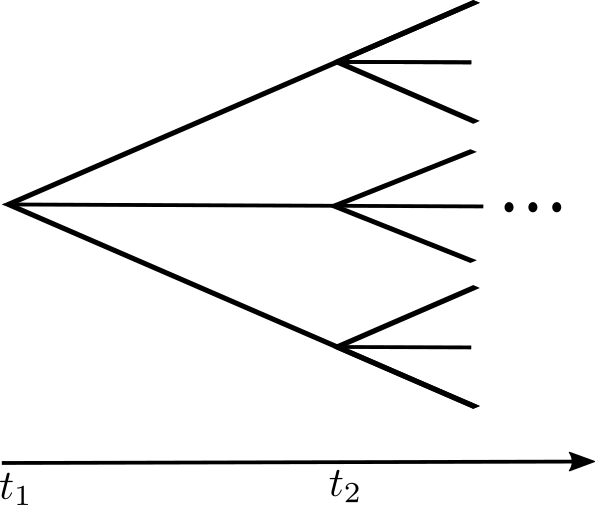}
  \caption{Branching structure of the quantum state in the decoherent histories approach.}
  \label{fig:CH_Psi}
\end{figure}

In the decoherent histories formulation, there is some lack of clarity about the class operator $C_{\alpha}$. It is presented as a mapping between the initial and final times $t_{1}$ and $t_{N}$. However, each projector in $C_{\alpha}$ acts at a different time in the sequence, such that each time enters into the theory as a distinct degree of freedom and the wavefunction in this theory must be a tensor product of states in the subspaces of $\mathcal{H}_{H}$:

\begin{equation}
    \left|\Psi_{U};\psi_{1}\left(t_{1}\right)\right\rangle =\left|\Psi\left(t_{N}\right)\right\rangle \otimes\ldots\otimes\left|\Psi\left(t_{2}\right)\right\rangle \otimes\left|\psi_{1}\right\rangle 
\end{equation}

This is the minimal structure of time-extended wavefunctions implied by the decoherent histories formalism. Each history is a structure appearing in the universal wavefunction, which is depicted as a branching tree in Fig. \ref{fig:CH_Psi}, where the branching events at initial time $t_{1}$ and the second time $t_{2}$ in the branching sequence could correspond, for example, to the three-dimensional macroscopic state space studied in Ref. \cite{strasberg2024first}. The inclusion of an initial condition $\left|\psi_{1}\right\rangle$ at $t_1$ is a constraint on the universal wavefunction $\left|\Psi_{U}\right\rangle$, which must be consistent with this constraint.

In the decoherent histories formalism, classical histories are defined in analogous fashion to the `present' in special relativity - the states $\left|\alpha_{k}\right\rangle$ are `actualized' only in an indexical way, for an observer who happens to exist at time $t_k$. However, whereas no observer can witness the entire branching structure defining $\left|\Psi_{U}\right\rangle$, they are committed to its existence unless the collapse of the wavefunction is explicitly included in the formalism. 

Cotler and Wilczek’s entangled-histories formalism \cite{7B1} can be viewed as a generalization the decoherent histories approach: instead of treating each consistent sequence of projectors as a separate classical alternative, it embeds every multi-time record into a single `history Hilbert space', so that superpositions of whole trajectories—and therefore genuine temporal entanglement—become legitimate quantum states. In this enlarged space the usual consistency condition reappears simply as orthogonality between history states, ensuring that the original probability assignments are recovered when one restricts to a decoherent family. The same construction is isomorphic to the MTS formalism \cite{nowakowski2018entangled}: with an appropriate inner product, every history state corresponds to a sequence of pre- and post-selected states. Hence, entangled histories not only subsume the standard  decoherent histories scheme but also furnish a unifying algebraic bridge to time-symmetric quantum mechanics, bringing the full toolkit of entanglement theory to the study of quantum processes in time.

\subsection{Two-State Vector Formalism and Multiple-Time Extensions}\label{sec:TSVF}

In parallel with the decoherent histories approach, wavefunctions in tensor products of time-localized Hilbert spaces feature in the time-symmetric approach to quantum mechanics pioneered by Aharonov et al. \cite{aharonov_time_1964}. This approach, which became the two state vector formalism (TSVF) \cite{aharonov_two-state_2008} and its multiple time state (MTS) generalization \cite{aharonov_multiple-time_2009,aharonov_each_2014}, treats quantum measurements which include dynamical boundary conditions on past and future times symmetrically.

The TSVF considers a system defined between preselection and postselection measurements at the times $t_{1}$ and $t_{2}$, respectively. The preselected state $\left|\psi\left(t_{1}\right)\right\rangle$ then travels forwards in time across the interval $\left[t_{1},t_{2}\right]$ in accordance with the TDSE, and the postselected state is represented by a vector in the conjugate space $\left\langle \phi\left(t_{2}\right)\right|$ which propagates backwards across the same time interval. The two oppositely orientated parts of the system can then be combined into a single `two state vector'

\begin{equation}\label{eq:TSV}
    \left\langle \phi\left(t_{2}\right)\right|\otimes\left|\psi\left(t_{1}\right)\right\rangle, 
\end{equation}

which exists in the composite Hilbert space constructed from distinct time-localized `universes' existing at single times \cite{aharonov_each_2014}

\begin{equation}
    \mathcal{H}_{t_{2}}^{\dagger}\otimes\mathcal{H}_{t_{1}}
\end{equation}

States in this Hilbert space are fundamentally (i) time nonlocal objects and (ii) built out of parts with opposite time orientations. Therefore, one solution to the apparent asymmetry under time reversal in quantum mechanics is to revise the notion of a quantum state itself to include two time degrees of freedom \cite{reznik1995time}. 

According to the TSVF, to obtain the probability of measuring the system in some state $\left|n\right\rangle$ at the intermediate time $t \in \left[t_{1},t_{2}\right]$, the system is propagated in \emph{both} time directions, from $t_{1}\rightarrow t$ and $t_{2}\rightarrow t$, such that the amplitude of the $n$-th outcome is given by sandwiching this state between the forwards and backwards-oriented parts of Eq. \ref{eq:TSV}:

\begin{equation}
    \left\langle \phi\left(t_{2}\right)\right|U\left(t_{2},t\right)\left|n\right\rangle \left\langle n\right|U\left(t,t_{1}\right)\left|\psi\left(t_{1}\right)\right\rangle 
\end{equation}

Then, assuming the Born rule, the normalized modulus-square of this amplitude yields the probability to obtain outcome $n$:

\begin{equation} \label{eq:ABL}
    P_{n}=\frac{\left|\left\langle \phi\left(t_{2}\right)\right|U\left(t_{2},t\right)\left|n\right\rangle \left\langle n\right|U\left(t,t_{1}\right)\left|\psi\left(t_{1}\right)\right\rangle \right|^{2}}{\underset{k}{\sum}\left|\left\langle \phi\left(t_{2}\right)\right|U\left(t_{2},t\right)\left|k\right\rangle \left\langle k\right|U\left(t,t_{1}\right)\left|\psi\left(t_{1}\right)\right\rangle \right|^{2}}
\end{equation}

This is the Aharonov-Bergmann-Lebowitz (ABL) probability rule. In addition, the theory of weak measurement values treats observables $\hat{O}$ that are weakly coupled to the state at $t$ to give the so-called weak value 

\begin{equation}
   \left\langle O\right\rangle _{w}\equiv\frac{\left\langle \phi\left(t_{2}\right)\right|U\left(t_{2},t\right)\hat{O}U\left(t,t_{1}\right)\left|\psi\left(t_{1}\right)\right\rangle}{\left\langle \phi\left(t_{2}\right)\right|U\left(t_{2},t_{1}\right)\left|\psi\left(t_{1}\right)\right\rangle} 
\end{equation}

The theory of weak values is developed at length in Refs. \cite{aharonov_how_1988,tamir_introduction_2013,aharonov_foundations_2014}, and has lead to many verified experimental predictions \cite{lundeen_experimental_2009,jordan_technical_2014,dressel2014colloquium}, lending weight to the idea that, in the quantum theories, the past and future affect each other symmetrically \cite{aharonov_can_2015}. It is a matter of current debate as to whether all phenomena predicted using the weak measurement protocol can be equally well captured by strong measurement protocols which are allowed to apply post-selection \cite{vaidman_past_2013,vaidman2017weak,hance2023weak}. However, there are cases where weak coupling (weak measurement) followed by post-selection lead to particularly interesting results \cite{hosoya2010strange,magana2014amplification}. For instance, a recent work demonstrates that the weak value measurement protocol (combining both weak coupling and post-selection) has a provable advantage in learning an unknown operator under the presence of certain types of noise \cite{schwartzman2024weak}. 

\begin{figure}
  \includegraphics[width=\linewidth]{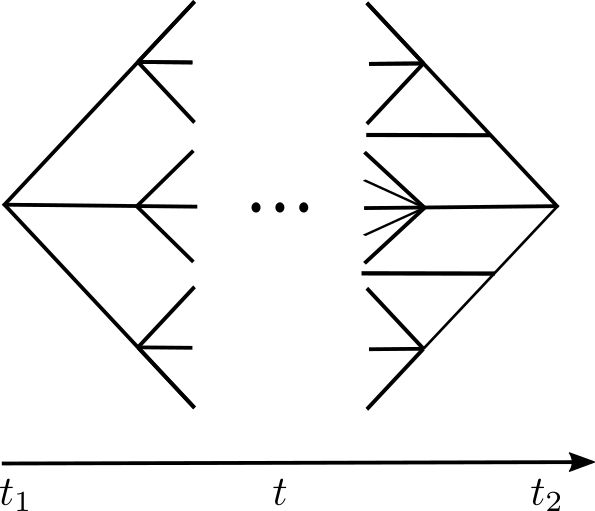}
  \caption{Branching structure of the quantum two-state according to the TSVF.}
  \label{fig:TSVF_Psi}
\end{figure}

The branching structure of the two-state vector in time is shown in Fig. \ref{fig:TSVF_Psi}. In addition to the forward-branching process initialized at time $t_{1}$, there is now a backwards-branching process running from the postselected state at $t_{2}$. The two processes meet at the measurement time $t$ \cite{,aharonov_two-state_2008,bopp2019bi,bopp2020avoid}. Note that, as indicated in Fig. \ref{fig:TSVF_Psi} the future- and past-directed parts of the two-state vector have different branching times and branch numbers, so the global two-state vector is not, in general, symmetric around the intermediate measurement time $t$. 

The TSVF opens up the intriguing possibility of states built up out of moments of time, each of which acts as a whole `universe' for quantum systems to explore. However, given a realist interpretation of the wavefunction, one may legitimately ask why the state of the system \emph{only} has a forward-directed component at $t_{1}$ and \emph{only} a backwards-directed component at $t_{2}$? If event symmetry is true, one expects there to be a backwards-directed component at $t_{1}$ and a forwards-directed part of the wavefunction at $t_{2}$. In other words, the physical state at the intermediate time $t$ could be considered as a \emph{source} boundary condition for propagation into the future and past, as well as a \emph{sink} for propagation from the pre- and post-selected boundary conditions at $t_{1}$ and $t_{2}$. In the TSVF, the state at time $t$ acts only as a sink. This type of propagation was described as `boundary-to-fixed point' (BTFP) in Ref. \cite{ridley2025time}. The converse type of two-time propagation is called `fixed-point-to-boundary' (FPTB).

The TSVF philosophy has been extended recently to states defined at multiple times \cite{aharonov_multiple-time_2009}, which was demonstrated to be mathematically equivalent to the process matrix formalism in Ref. \cite{silva_connecting_2017}. This multiple-time-state (MTS) approach models sequences of pre- and post-selections in terms of tensor products of bra and ket vectors, respectively. As such, it does not diverge from the TSVF insofar as it employs an identical notion of what two-time evolution is. Rather, the MTS simply involves cases where the time domain for an experiment is separated into more than one time segment, divided by more than two boundary conditions. It is a versatile formalism which can capture many situations inaccessible to standard quantum mechanics. An interesting case, considered in Ref. \cite{aharonov_multiple-time_2009}, that can be captured within MTS is akin to entanglement between the forward- and backward evolving states

\begin{equation}
    \underset{i}{\sum}\left\langle i\left(t_{2}\right)\right|\otimes\left|i\left(t_{1}\right)\right\rangle.
\end{equation}

This state describes a a situation in which the forward-evolving component of the wavefunction at time $t_{1}$ carries information which is perfectly correlated with the backward-evolving component at the later time $t_{2}$. It can be visualized as a closed time loop. We can take this idea further in the MTS, constructing a Hilbert space of \emph{four} times

\begin{equation}
    \mathcal{H}^{\dagger}_{t_{4}}\otimes\mathcal{H}_{t_{3}}\otimes\mathcal{H}^{\dagger}_{t_{2}}\otimes\mathcal{H}_{t_{1}},
\end{equation}

in which an unusual, massively entangled state can be defined, i.e.

\begin{equation}\label{eq:i_4_Entangled}
    \underset{i}{\sum}\left\langle i\left(t_{4}\right)\right|\otimes\left|i\left(t_{3}\right)\right\rangle\otimes\left\langle i\left(t_{2}\right)\right|\otimes\left|i\left(t_{1}\right)\right\rangle.
\end{equation}

The state defined in Eq. \eqref{eq:i_4_Entangled} captures a situation in which the process of `circulation' itself in the two time segments $\left[t_{1},t_{2}\right]$, $\left[t_{3},t_{4}\right]$ is completely entangled. Thus entanglement between entire time-extended processes can be captured within the MTS formalism.

Following on from the MTS structure, Aharonov \emph{et al.} \cite{aharonov_each_2014} have coined the phrase `each instant of time a new universe' (ETNU) to describe the wavefunction across regions of time, constructed from wavefunctions defined at the boundaries of tiny time `bricks'. In this formulation, at each time, two Hilbert spaces are defined, one for each time direction. If time is discretized into $N_{t}-1$ `bricks', we can assign a new Hilbert space to the pair of times at each `brick' boundary, i.e. the total Hilbert space, in their formulation, has the structure:

\begin{equation}
    \mathcal{H}_{N_{t}}^{\dagger}\otimes\mathcal{H}_{N_{t}-1}\otimes...\otimes\mathcal{H}_{t_{1}}^{\dagger}\otimes\mathcal{H}_{t_{1}},
\end{equation}

such that the global quantum state possesses $2N_{t}-1$ temporal dimensions. 

\begin{figure}
  \includegraphics[width=\linewidth]{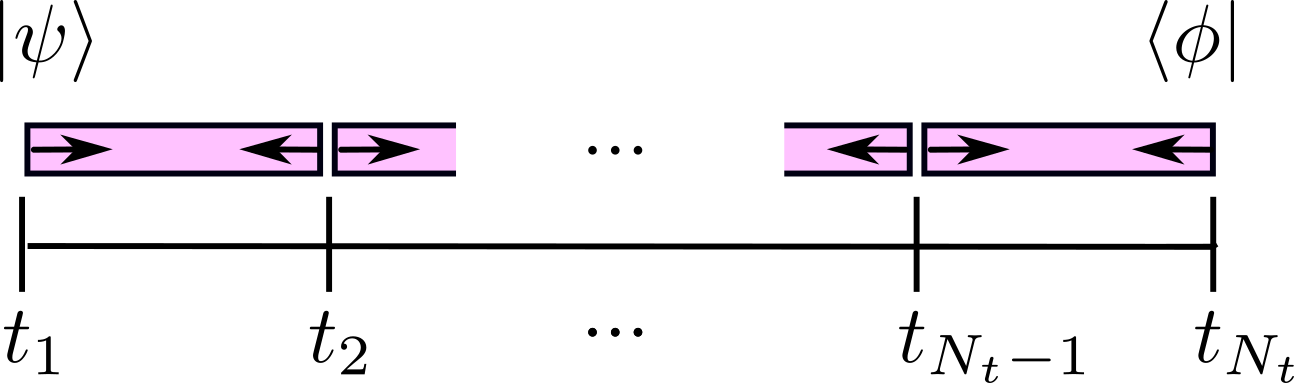}
  \caption{The sequence of time `bricks' with boundary-to-fixed-point (BTFP) propagation that characterizes the ETNU formulation.}
  \label{fig:ETNU}
\end{figure}

In Fig. \ref{fig:ETNU}, we display a schematic representation of the ETNU time propagation between a pre-selection measurement of the state $\left|\psi\right\rangle$ at $t_{1}$ and a post-selection, fixing the state to $\left\langle\phi\right|$ at $t_{N_{t}}$. The `bricks' which make up the temporal region $t_{1},t_{N_{t}}$ are shown in pink in Fig. \ref{fig:ETNU}, where the arrows indicate a sequence of tiny FPTB-style time evolutions. Also included in the ETNU formalism is a novel conception of time evolution itself - the standard dynamical picture of the TSVF is replaced with \emph{correlations} between adjacent states in the sequence:

\begin{equation}
    \left\langle \phi\right|\underset{i}{\sum}U_{t_{N_{t}}t_{N_{t}-1}}\left|i\right\rangle _{t_{N_{t}}}\left\langle i\right|_{t_{N_{t}-1}}\ldots\underset{i}{\sum}U_{t_{2}t_{1}}\left|i\right\rangle _{t_{2}}\left\langle i\right|_{t_{1}}\left|\psi\right\rangle,
\end{equation}

where $U_{t_{2}t_{1}}$ is numerically equal to the unitary evolution operator $U\left(t_{2},t_{1}\right)$. \cite{aharonov_each_2014} then `contract' the vectors belonging to the past and future boundary conditions, i.e. they allow the bras and kets in the above expression to simply overlap, generating the amplitudes of large temporal sequences of measurement events. The modulus-squared of these amplitudes is then set equal to the probability of the sequence, assuming the Born rule.

\subsection{Transactional Interpretation}

One of the most striking extant retrocausal formulations of quantum mechanics can be found in the transactional interpretation (TI) of Cramer \cite{cramer1980generalized,cramer_transactional_1986,cramer_overview_1988}. The TI models every physical process in terms of a `transaction' composed of both the standard (forwards time) solutions to the Schr{\"o}dinger equation (the `offer' wave) and their backwards time complex conjugates (the `confirmation' wave). In this theory, quantum processes are modelled in terms of physical waves emanating from an emitter to an absorber. The emitter sends out the retarded wave in the forwards time direction, and also an advanced wave in the direction of negative time. The absorber exists at a later time to the emitter, and it also emits both retarded and advanced waves. In a quantum transaction, retarded and advanced waves are exchanged across the same region of time, transferring energy, momentum and angular momentum.

The usual quantum state propagating from the emitter to the $n$-th absorber is denoted as a weighted `ket' vector $\langle\psi_{n}|\Psi\rangle|\psi_n\rangle$. This is combined with the advanced response, denoted with a `bra' vector $\langle\Psi|\psi_{n}\rangle\langle\psi_{n}|$, from the absorber, called the `confirmation' wave \cite{kastner2012possibilist,kastner_transactional_2013,kastner2016transactional}. This two-way process between emitters and absorbers occurs across entire time intervals and models the irreversible reduction of the state into one of the outcomes $|\psi_{n}\rangle$ with a transaction amplitude, or `echo' given by

\begin{equation}
   \langle\psi_{n}|\Psi\rangle\langle\Psi|\psi_{n}\rangle. 
\end{equation}

This quantity is equal to the Born measure. Thus, in the TI, the quantum probability appears as the scaling amplitude for a process involving a bidirectional transaction between emitter and absorber, playing a physical role analogous to the intensity of an electromagnetic wave \cite{cramer_transactional_1986,kastner_transactional_2013}.

In this connection, it is important to stress that the roots of the original TI lie in the time-symmetric Wheeler-Feynman formulation of classical electrodynamics \cite{wheeler1945interaction,wheeler1949classical}, which was later extended by Davies to the quantum electrodynamical case using an S-matrix approach \cite{davies1970quantum,davies1971extension,davies1972extension}. The Wheeler-Feynman theory described radiative processes in terms of a combination of advanced and retarded waves. Such processes occur in spacetime, and their description is Lorentz-covariant. Cramer emphasizes that so too is the TI, because it describes processes across regions of spacetime, \emph{not} temporal regions \cite{cramer_transactional_1986}. Also, the TI makes use of both positive and negative energy/time solutions to the Klein-Gordon equation in the non-relativistic limit, so relativity is built into this theory from the outset \cite{kastner2012possibilist}.

\subsection{Fixed-point formulation}

The pseudo-spin extension of Bauer adds a second temporal degree of freedom for each moment on the regular timeline. Another recent approach which does this is the fixed-point formulation (FPF) developed in Refs. \cite{ridley2023quantum,ridley2025time}. The FPF replaces the usual single-time wavefunction of quantum mechanics with a multiple-time ontology, specified on the Keldysh contour shown in Fig. \ref{fig:Time_contour}. The FPF, moreover, provides a novel axiomatization of quantum theory in which the Born rule can be \emph{derived} from the core postulates of the theory.

The FPF implements the ETNU philosophy \emph{at every point on the Keldysh contour}, i.e. the state space of the system is replicated at each time \emph{and} at each time orientation. This is distinct from the multiple-time state formalism, where backwards-oriented parts of the wavefunction at time $t$ exist in the dual Hilbert space $\mathcal{H}_{t}^{\dagger}$. In the FPF, a distinct Hilbert space is assigned to each contour time, $\mathcal{H}_{t_{i}}^{\alpha}$, where $\alpha\in\left\{ f,b\right\}$ denotes the upper or lower branch of $\gamma$. The resulting enlarged Hilbert space $\mathcal{H}_{\gamma}$ is called the \emph{Contour space}:

\begin{gather}\label{event_space}
    \mathcal{H}_{\gamma}=\mathbb{C}\oplus\overset{\infty}{\underset{N_{t}=1}{\bigoplus}}\overset{N_{t}}{\underset{i=1}{\bigotimes}}\int_{\gamma}^{\oplus}\mathcal{H}_{z_{i}}\\
    =\mathbb{C}\oplus \int_{\gamma}^{\oplus}\mathcal{H}_{z_{1}}\oplus\int_{\gamma}^{\oplus}\mathcal{H}_{z_{1}}\otimes\mathcal{H}_{z_{2}}\oplus\ldots
\end{gather}

States defined in $\mathcal{H}_{\gamma}$ are superpositions of all possible time sequences on the Keldysh contour. Note that, for a relativistically covariant extension of this idea, one should consider a space that allows for all possible \emph{orderings} of the densely-packed times on $\gamma$. This would allow for translations between reference frames in which the ordering of times is permuted. This generalization of the FPF could capture processes with in a superposition of time orderings, much like the indefinite causal ordering seen in recent experimental \cite{rubino2017experimental,goswami2018indefinite,rozema2024experimental,stromberg2024experimental,guo2024experimental} and theoretical \cite{ebler2018enhanced,felce2020quantum,zhao2020quantum,chiribella2021symmetries,chiribella2022quantum,liu2025tsirelson} works. For the purposes of this work, it is sufficient to consider time sequences with fixed ordering, in the non-relativistic scenario. 

The FPF embeds two-time Keldysh dynamics into the ontology of the wavefunction itself. This is done in two steps: first, the universal quantum state is specified as a member of $\mathcal{H}_{\gamma}$, i.e, we set 

\begin{equation}\label{eq:Psi_U}
    \left|\Psi_{U}\right\rangle=\left|0\right\rangle+\overset{\infty}{\underset{N_{t}=1}{\Sigma}}\overset{N_{t}}{\underset{i=1}{\otimes}}\int_{C}\left|\psi_{i}\right\rangle dz_{i}
\end{equation}

as the universal state, corresponding to the usual ontological postulate of textbook quantum mechanics (systems are represented by vectors in Hilbert spaces). We may refer to this as the `time-full' representation, by way of contrast with the `timeless' universal wavefunction in the PW approach. 

Second, we connect oppositely-oriented parts of the wavefunction independently on $\gamma_{f}$ and $\gamma_{b}$ by specifying a contour-time-dependent time derivative everywhere on $\gamma$

\begin{equation}
i\hbar\partial_{t^{\alpha}}\left|\Psi^{\alpha}\left(t^{\alpha}\right)\right\rangle =H^{\alpha}\left(t^{\alpha}\right)\left|\Psi^{\alpha}\left(t^{\alpha}\right)\right\rangle .
\end{equation}

This corresponds to the second core postulate found in textbooks (states evolve unitarily in time in accordance with the time-dependent Schr{\"o}dinger equation), and generates the usual unitary mappings between time-localized Hilbert spaces.

We now formalize the notion of a \emph{fixed point} in Contour Space, which corresponds to a temporal boundary condition on the Keldysh contour:

\textit{A fixed point at time $t$ is a temporal part of the wavefunction in the $\mathcal{H}_{t}^{b}\otimes\mathcal{H}_{t}^{f}$ subspace, with equal $f$ and $b$ parts.}

\begin{figure}
  \includegraphics[width=.85\linewidth]{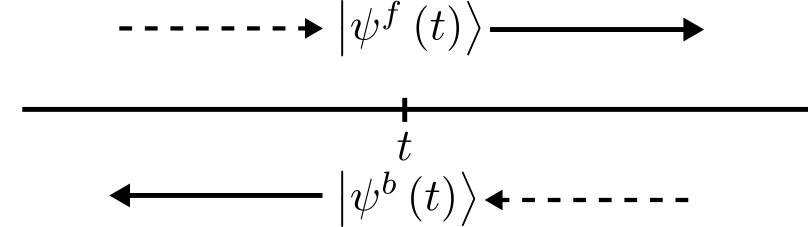}
  \caption{A single fixed point on the Keldysh contour.}
  \label{fig:1FP}
\end{figure}
The concept of a fixed point as both a source and sink for future- and past-directed processes is illstrated schematically in Fig. \ref{fig:1FP}. We also introduce the special notation
\begin{equation} \label{eq:1FP}
    \llbracket\psi\rrbracket_{t} \equiv | \psi^{b} \left(t^{b}\right)\rangle \otimes| \psi^{f} \left(t^{f}\right)\rangle 
\end{equation}
for a fixed point specifying the state as $| \psi \left(t\right)\rangle$ at time $t$. Physically, this corresponds to an event specifying the state at $t$, or to a time-indexed projection in the decoherent histories language. As discussed in Section \ref{sec:Bauer}, we may think of the `present' time $t$ as `pinched' in between the upper-branch and lower-branch times $t^{f}$, $t^{b}$. For a full description of a measurement connecting a preparation at some initial time $t_{1}$ to an observation of the state at the later time $t_{2}$ across the region $\left[t_{1},t_{2}\right]$, at least two fixed points must be present from the atemporal perspective, i.e. we project Eq. \eqref{eq:Psi_U} onto its component in the reduced Hilbert space $\mathcal{H}_{\gamma}\left(N_{t}\right)$:
\begin{equation}
  \mathcal{H}_{\gamma}\left(N_{t}\right)=\mathcal{H}_{t_{N_{t}}}^{b}\otimes\mathcal{H}_{t_{N_{t}}}^{f}\otimes\ldots\otimes\mathcal{H}_{t_{1}}^{b}\otimes\mathcal{H}_{t_{1}}^{f}
  \label{eqn:Hilbert_Space}
\end{equation}

A quantum history sequence is then defined in terms of products of fixed points:

\textit{A quantum history $h_{\mathbf{k}}$ extending across the time range $\left[t_{1},t_{2}\right]$ is a product state constructed from a sequence ${\mathbf{k}}=\left\langle k_{1},...,k_{N_{t}}\right\rangle $ of $N_{t}\geq2$ fixed points
\begin{equation}\label{eq:Quantum_History}
    h_{\mathbf{k}}=\overset{N_{t}}{\underset{i=1}{\bigotimes}}\left\llbracket \psi_{k_{i}}\right\rrbracket _{t_{i}}
\end{equation}
connected by unitary mappings and bounded by fixed points at $t_{1}$ and $t_{2}$.}

In Eq. (\ref{eq:Quantum_History}), each $k_{i}$ in a history $h_{\mathbf{k}}$ ranges over a complete basis set spanning $\mathcal{H}_{t_{i}}^{\alpha}$. To allow us to apply rules of probabilistic reasoning to such histories, we define a \emph{family} of quantum histories $\mathcal{F}_{H}$ by imposing the consistency condition that any pair of histories in a family $\left\{ \left|h_{\mathbf{k}}\right\rangle \right\} $ must be non-overlapping:
\begin{equation}\label{eq:Quantum_History_Consistency}
    \left\langle h_{\mathbf{l}}\right.\left|h_{\mathbf{k}}\right\rangle =\delta_{\mathbf{kl}},
\end{equation}

where $\mathbf{k}\neq\mathbf{l}$ if $\left\llbracket \psi_{k_{i}}\right\rrbracket _{t_{i}}\neq\left\llbracket \psi_{l_{i}}\right\rrbracket _{t_{i}}$ for at least one value of $i \in \left[1,...,N_{t}\right]$. The consistency condition Eq. \eqref{eq:Quantum_History_Consistency} prevents the overlap of histories composed of different numbers of times $N_{t}$.

Following the terminology of Vaidman \cite{vaidman_schizophrenic_1998}, the \emph{measure of existence} of a history may now be defined as the relative size of the wavefunction region occupied by that history:

\textit{The measure of existence $m\left(h_{\textbf{k}}\right)$ of a quantum history $h_{\mathbf{k}}$ containing $N_{t}$ fixed points in the time range $\left[t_{1},t_{2}\right]$, is the ratio of the integral of the wavefunction $\triangle\Psi_{\textbf{k}}$ along this history, to that of all histories 
\begin{equation}\label{eq:MOE}
    m\left(h_{\textbf{k}}\right)=\frac{\ensuremath{\triangle\Psi_{\textbf{k}}}}{\underset{\mathbf{k'}}{\sum}\ensuremath{\triangle\Psi_{\textbf{k'}}}}
\end{equation}
consistent with the fixed point boundary conditions.
}

Fixed point boundary conditions are imposed by taking the inner product of the integrated wavefunction with the `sink' state defined at the upper limits of the $2(N_{t}-1)$ segment integrals.  

To complete its reformulation of quantum mechanics, the FPF then makes the following statistical postulate, called the \emph{Vaidman rule}, which corresponds to the usual measurement postulate in textbook quantum theory.

\textit{The quantum probability of a quantum history is equal to its measure of existence in the universal wavefunction.} 

No explicit mathematical expression has been assumed for the measure of existence. The \emph{mathematical} form of the Born measure, its extension to the ABL rule when post-selection measurements are carried out, and the general formula for the probability of an $N_{t}$-time history can all be \emph{derived} from the Vaidman rule and the structural postulates of the FPF \cite{ridley2023quantum,ridley2025MRW}. 

To see how histories emerge from the universal wavefunction in practice, consider a sequence of $N_{t}$ strong measurements in the bases $\left\{ \left|k_{1}\right\rangle \right\} $, $\left\{ \left|k_{2}\right\rangle \right\} $,...,$\left\{ \left|k_{N_t}\right\rangle \right\} $, defining a family $\mathcal{F}$ of histories. Setting

\begin{equation}
    \left|\Psi^{\alpha}\left(t_{i}^{\alpha}\right)\right\rangle =\underset{k_{i}}{\sum}c_{k_{i}}^{\alpha}\left|k_{i}\right\rangle,
\end{equation}

one can expand the universal state vector 
$\left|\Psi_{U}\right\rangle $ defined in Eq. $\mathcal{H}_{\gamma}$ in this basis:

\begin{equation}\label{eq:Psi_Universe_Expanded}
    \left|\Psi_{U}\right\rangle =\bigotimes_{i=1}^{N_{t}}\underset{k_{i},l_{i}}{\sum}c_{k_{i}}^{b}c_{l_{i}}^{f}\left|k_{i}\right\rangle \left|l_{i}\right\rangle 
\end{equation}

Recall that $\left|\Psi_{U}\right\rangle $ exists in the reduced contour space $\mathcal{H}_{C}\left(N_{t}\right)$, and from Eq. \eqref{eq:Psi_Universe_Expanded} we see that the upper- and lower-branch parts of $\left|\Psi_{U}\right\rangle $ at the contour times $t^{f}_{i}$, $t^{b}_{i}$ are in general \emph{not} equal. Nevertheless, when one carries out a strong measurement at time $t_{i}$, the observer \emph{constrains} the region of $\left|\Psi_{U}\right\rangle $ in which they are carrying out the observation, imposing the condition that the upper-and lower-branch components of the wavefunction must be identical at this time. This reflects the intuition, also expressed in Ref. \cite{aharonov_each_2014}, that a strong measurement serves as \emph{both} a pre- and post-selection for the future and past, respectively. So there is a fixed point at each cross-branch time-slice of $\left|\Psi_{U}\right\rangle $, a condition which can be imposed by projecting Eq. \eqref{eq:Psi_Universe_Expanded} along the family of histories with the following projection operator:

\begin{equation}\label{eq:Projection_F}
    \hat{\mathbf{P}}_{\mathcal{F}}\equiv\bigotimes_{i=1}^{N_{t}}\underset{k_{i}}{\sum}\left|k_{i}\right\rangle \left|k_{i}\right\rangle \left\langle k_{i}\right|\left\langle k_{i}\right|=\bigotimes_{i=1}^{N_{t}}\underset{k_{i}}{\sum}\left\llbracket k_{i}\right\rrbracket \left\llbracket k_{i}\right\rrbracket ^{\dagger}
\end{equation}

This procedure results in

\begin{align}
\hat{\mathbf{P}}_{\mathcal{F}}\left|\Psi_{U}\right\rangle &=\bigotimes_{i=1}^{N_{t}}\underset{k_{i}}{\sum}c_{k_{i}}^{b}c_{k_{i}}^{f}\left\llbracket k_{i}\right\rrbracket\\
&=\underset{k_{1},...,k_{N_{t}}}{\sum}c_{k_{1}}^{b}c_{k_{1}}^{f}...c_{k_{N_{t}}}^{b}c_{k_{N_{t}}}^{f}\left\llbracket k_{N_{t}}\right\rrbracket \otimes...\otimes\left\llbracket k_{1}\right\rrbracket \label{eq:History_superposition},
\end{align}

i.e. the imposition of the measurement context restricts the observer to the part of the universal wavefunction which is a superposition of quantum histories. This aligns with the prescription of Adlam in Ref. \cite{adlam2022two} that `all-at-once' retrocausal models, also termed `block instantiation' models, should be construed in terms of \emph{constraints}, specifying which histories one can distribute probabilities over. 

Next, one can impose boundary conditions by only distributing over those histories in which the boundary conditions are satisfied. For instance, in the simple case of a three-point measurement involving the times $t_{1}$, $t$ and $t_{2}$, there is typically a \emph{preparation} of the state at some definite value $\left|\psi\left(t_{1}\right)\right\rangle$ at time $t_{1}$, followed by a measurement of $\left|\phi\left(t_{2}\right)\right\rangle$ at time $t_{2}$. Without loss of generality, the prepared and post-selected states are taken to be members of complete bases $\left|\psi\right\rangle = \left|\psi_{i}\right\rangle \in \left\{ \left|\psi_{k}\right\rangle\right\}$, $\left|\phi\right\rangle = \left|\phi_{i}\right\rangle \in \left\{ \left|\phi_{k}\right\rangle\right\}$. The measurement then defines a family of three-fixed-point histories 

\begin{equation}
\mathcal{F}_{3}:\left\llbracket \psi\right\rrbracket _{t_{1}}\otimes\left\{ \left\llbracket a_{i}\right\rrbracket _{t}\right\}\otimes\left\llbracket \phi\right\rrbracket _{t_{2}}
\label{eqn:3FP_Family}
\end{equation}

whose measure of existence in $\left|\Psi_{U}\right\rangle$ can be evaluated explicitly, using the procedure in Ref. \cite{ridley2023quantum}. 

\begin{figure}
  \includegraphics[width=0.6\linewidth]{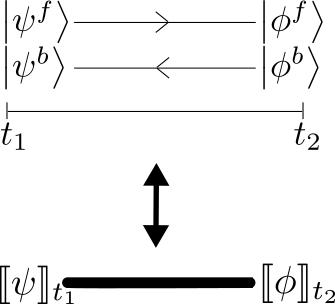}
  \caption{A compact representation of a two-time channel connecting two fixed points (a history segment) in the FPF.}
  \label{fig:history_segment}
\end{figure}

\begin{figure}
  \includegraphics[width=\linewidth]{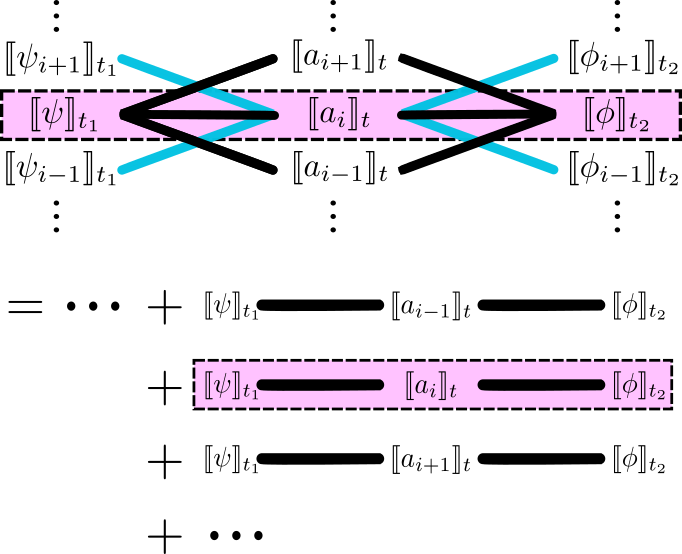}
  \caption{Decomposition of the wavefunction for a measurement at $t$ with pre- and post-selection at $t_{1}$, $t_{2}$ into the family of histories $\mathcal{F}_{3}$.}
  \label{fig:ABL_Compact}
\end{figure}

We can introduce a \emph{history segment} as a tensor product of two fixed points on the Keldysh contour, connected by two time channels on the $f$ and $b$ branches. An arbitrary history segment is depicted schematically in Fig. \ref{fig:history_segment}, where we also introduce a compact thick-line representation for the two-channel history for ease of notation. 
The histories in $\mathcal{F}_{3}$ that are consistent with the pre- and post-selection constraints and with the measurement basis are illustrated schematically with black lines in Fig. \ref{fig:ABL_Compact}. The pink box indicates the member of this history set for which the probability is evaluated. Those history segments which are consistent with the measurement context, but not the specific constraints imposed by the pre- and post-selection, are represented with blue lines. The blue-line processes constitute a part of the universal wavefunction that is `cut away' when the observer carries out the pre- and post-selection, as mathematically described above by the projection operator in Eq. \eqref{eq:Projection_F}. This constrained wavefunction is mathematically equivalent to a superposition of three-fixed-point histories, indicated diagrammatically in the right-hand side of the equation in Fig. \ref{fig:ABL_Compact}, reflecting the form of Eq. \eqref{eq:History_superposition}. As such, only the black line histories are included in the evaluation of the measure of existence from Eq. \eqref{eq:MOE}, which turns out to be equal to the ABL measure in Eq. \eqref{eq:ABL} \cite{ridley2023quantum,ridley2025MRW}.

Finally, we note that a time operator can be introduced for a single fixed point, based on an `unfolding' of the time operator in the Bauer formalism:

\begin{equation}\label{eq:t_FP}
    \hat{t}^{FP}=\frac{i}{2}\left[\frac{dD^{\dagger}}{d\varepsilon}\otimes\mathbf{I}^{f}+\mathbf{I}^{b}\otimes\frac{dD}{d\varepsilon}\right]|_{\varepsilon=0}
\end{equation}

In Section \ref{sec:Conceptual}, we show how the action of this operator on a single fixed point returns the observable time. 

The operator $\hat{t}^{FP}$ is a linear combination of the corresponding time operators on the upper and lower branches of the Keldysh contour. Intuitively, the time at which a fixed point exists is `pinched' in-between times with opposite orientations. This, we suggest, tells us something about the nature of time itself, something which is not contained in the operationalist PW approach. Every instant of time seems to disappear in both directions - one cannot specify the present without simultaneously specifying how it slips away into the past and future.

We cannot, however, introduce a time operator for the universal wavefunction or for quantum histories - as intrinsically multiple-time objects, asking for `the' temporal location of the universe or a manifestly time-extended history segment is meaningless. This is analogous to the situation described in Section \ref{sec:PW} for the PW approach - the universal wavefunction is both `timeless' and `time-full', depending on the perspective of the (hypothetical) observer. 

\section{Connections between the approaches}\label{sec:Comparison}

\begin{figure}[htp]

  \begin{subfigure}{\linewidth}
    \subcaption{One-way} \label{fig:one-way}
    \includegraphics[clip, width=\linewidth]{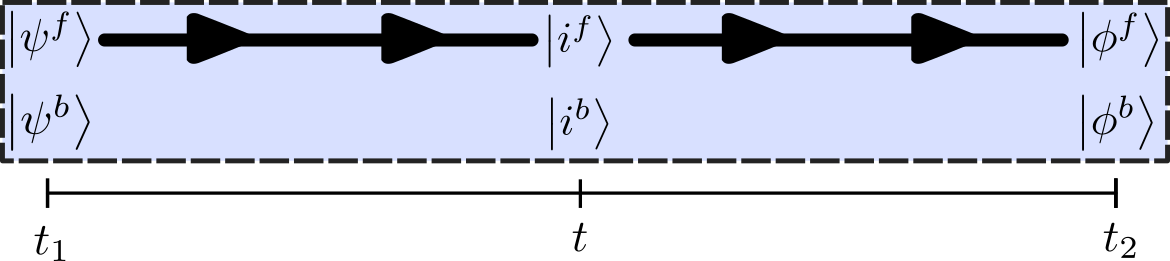}
  \end{subfigure}%

  \begin{subfigure}{\linewidth}
    \subcaption{Corridor} \label{fig:corridor}
    \includegraphics[clip, width=\linewidth]{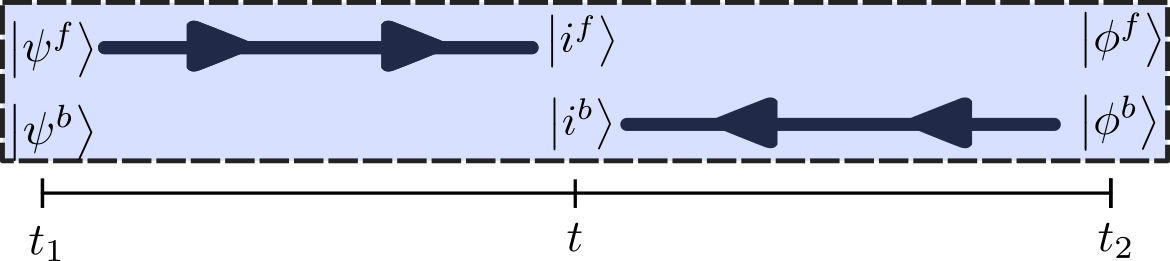}
  \end{subfigure}%
  
  \begin{subfigure}{\linewidth}
    \subcaption{Motorway} \label{fig:motorway}
    \includegraphics[clip, width=\linewidth]{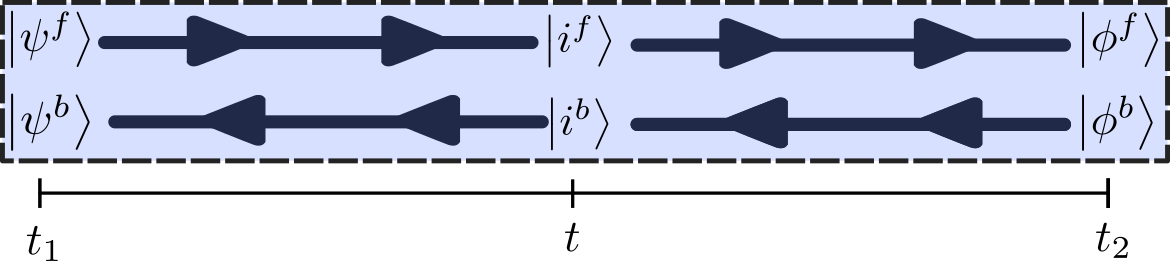}
  \end{subfigure}
\caption{Schematic representation of the regions and direction of time propagation between three consecutive boundary conditions considered within (a) the forward evolution in time given by the standard Schr{\"o}dinger/PW dynamics (it should be noted, however, that within the Schr{\"o}dinger picture time is an external, global parameter, while in the PW approach it is given by an operator), (b) the `corridor' description of TSVF/MTS which consists of forward/backward evolution from the initial/final boundary condition, respectively, (c) the `motorway' description of Bauer/FPF which results from the doubling of the Hilbert space, Hamiltonian spectrum and time evolution.} \label{fig:schematic}
\end{figure}

All of the formalisms considered in this work may be categorized in terms of the type of time propagation they allow. The different cases are illustrated schematically in Fig. \ref{fig:schematic}, where we show three different types of time propagation across two regions of time separated by three boundary conditions at $t_{1}$, $t$ and $t_{2}$. The intermediate time $t$ could, for instance, correspond to a strong measurement in the basis $\left\{ \left|i\right\rangle \right\}$, whereas the boundary conditions constraining the dynamics at $t_{1}$ and $t_{2}$ are imposed via pre- and post-selection. In Fig. \ref{fig:one-way}, the standard Schr{\"o}dinger evolution is shown. We refer to this as `one-way' evolution because the states are directed towards the future at all points in the temporal region $\left[t_{1},t_{2}\right]$. Fig. \ref{fig:corridor} illustrates time evolution where the state is future-directed in the past of $t$ but past-directed in its future, meeting at the middle in what we term a `corridor' propagation. The corridor structure is typically seen in the TSVF. Finally, Fig. \ref{fig:motorway} shows two-time propagation where at every point in $\left[t_{1},t_{2}\right]$, there exist two time orientations. This is called a `motorway' propagation, and is characteristic of both the Bauer approach and the FPF.

We now use this schematic to draw connections and distinctions between the different formalisms considered in this work, both at the mathematical and conceptual levels.

\subsection{Mathematical Connections}\label{sec:Mathematical}

We have so far compared the different theories in virtue of their structural aspects. Specifically, we have described the Hilbert spaces of each theory, the procedures for computing probabilities and the procedures for constructing a time operator. In this section we demonstrate a further mathematical connection between two of the theories considered in this work.  

First of all, we note that the Bauer and FPF formalisms can be shown to be isomorphic in the case of Hamiltonians with a continuous spectrum, as they both make use of the full Keldysh contour, i.e. the `motorway' two-time structure shown in Fig. \ref{fig:motorway}. Indeed, suppose we expand a generic spinor state $\left|\Psi\right\rangle$ in the Bauer space, Eq. \eqref{eq:Hilbert_Bauer}, in terms of the eigenspectrum defined by Eq. \eqref{eq:pseigenspectrum}:

\begin{equation}
    \left|\Psi\right\rangle =\overset{\infty}{\underset{0}{\int}}dE\underset{\alpha}{\sum}\left\langle E,\alpha\right.\left|\Psi\right\rangle \left|E,\alpha\right\rangle 
\end{equation}

As discussed in Section \ref{sec:Bauer}, we can use the projection operators $P$, $\bar{P}$ defined in Eqs. \eqref{eq:P} and \eqref{eq:I-P} to project $\left|\Psi\right\rangle$ along its forwards and backwards-directed components, which are then projected along the ket $\left|t\right\rangle$:

\begin{equation}
\begin{split}
    \left|\Psi^{f}\left(t\right)\right\rangle &=\left|P\Psi\left(t\right)\right\rangle \equiv\left\langle t\right|P\left|\Psi\right\rangle\\
    &=\overset{\infty}{\underset{0}{\int}}dE\left\langle E,f\right.\left|\Psi\right\rangle \left\langle t\right.\left|E,f\right\rangle  
\end{split}   
\end{equation}

\begin{equation}
\begin{split}
    \left|\Psi^{b}\left(t\right)\right\rangle &=\left|\bar{P}\Psi\left(t\right)\right\rangle \equiv\left\langle t\right|\bar{P}\left|\Psi\right\rangle \\
    &=\overset{\infty}{\underset{0}{\int}}dE\left\langle E,b\right.\left|\Psi\right\rangle \left\langle t\right.\left|E,b\right\rangle 
\end{split}
\end{equation}

This enables us to make a direct connection between the spinor states in Bauer's approach and the fixed point states in the FPF - the latter are simply an `unfolding' of the former:

\begin{equation}
\begin{split}
    \left\llbracket \Psi\right\rrbracket _{t} &\equiv \left|\Psi^{b}\left(t\right)\right\rangle \otimes\left|\Psi^{f}\left(t\right)\right\rangle \\ &=\left|\bar{P}\Psi\left(t\right)\right\rangle \otimes\left|P\Psi\left(t\right)\right\rangle 
\end{split}
\end{equation}

Since the time operator in Eq. \eqref{eq:t_diagonal} is given in the block-diagonal form, it has eigenstates which can be written in the spinor form:

\begin{equation}
    \left|t\right\rangle =\left[\begin{array}{c}
\left|t,f\right\rangle \\
0
\end{array}\right]+\left[\begin{array}{c}
0\\
\left|t,b\right\rangle 
\end{array}\right]
\end{equation}

The components of these spinors have the following overlaps with the corresponding energy components defined in Eq. \eqref{eq:pseigenspectrum}:

\begin{equation}
    \left\langle t,f\right.\left|E,f\right\rangle =\frac{e^{iEt}}{\sqrt{2\pi}}
\end{equation}
\begin{equation}
    \left\langle t,b\right.\left|E,b\right\rangle =\frac{e^{-iEt}}{\sqrt{2\pi}}
\end{equation}

\begin{widetext}

Using these identities, the definition of the time operator components (Eqs. \eqref{eq:t_f} and \eqref{eq:t_b}) and the energy shift operator (Eqs. \eqref{eq:D} and \eqref{eq:D^+}) we are able to calculate the action of $\hat{t}$ on the $f$ and $b$ components of $\left|\Psi\right\rangle$, in the $t$- representation. 
\begin{equation}
\begin{split}
    \left\langle t\right|\hat{t}\left|P\Psi\right\rangle &= \left(\left[\left\langle t,f\right|,0\right]+\left[0,\left\langle t,b\right|\right]\right)\left(\begin{array}{cc}
    \hat{t}^{f} & 0\\
    0 & \hat{t}^{b}
    \end{array}\right)\overset{\infty}{\underset{0}{\int}}dE\left[\begin{array}{c}
    \left|E,f\right\rangle \left\langle E,f\right.\left|\Psi\right\rangle \\
    0
    \end{array}\right]\\
    &= \overset{\infty}{\underset{0}{\int}}dE\left\langle t,f\right|\hat{t}^{f}\left|E,f\right\rangle \left\langle E,f\right.\left|\Psi\right\rangle\\
    &= i\overset{\infty}{\underset{0}{\int}}dE\frac{d}{d\varepsilon}\left.\left(\left\langle t,f\right.\left|E-\varepsilon,f\right\rangle \right)\right|_{\varepsilon=0}\left\langle E,f\right.\left|\Psi\right\rangle\\
    &= t\overset{\infty}{\underset{0}{\int}}dE\frac{e^{iEt}}{\sqrt{2\pi}}\left\langle E,f\right.\left|\Psi\right\rangle = t\left|\Psi^{f}\left(t\right)\right\rangle 
\end{split}
\end{equation}

\begin{equation}
\begin{split}
    \left\langle t\right|\hat{t}\left|\bar{P}\Psi\right\rangle &=\left(\left[\left\langle t,f\right|,0\right]+\left[0,\left\langle t,b\right|\right]\right)\left(\begin{array}{cc}
    \hat{t}^{f} & 0\\
    0 & \hat{t}^{b}
    \end{array}\right)\overset{\infty}{\underset{0}{\int}}dE\left[\begin{array}{c}
    0\\
    \left|E,b\right\rangle \left\langle E,b\right.\left|\Psi\right\rangle 
    \end{array}\right] \\
    &= \overset{\infty}{\underset{0}{\int}}dE\left\langle t,b\right|\hat{t}^{b}\left|E,b\right\rangle \left\langle E,b\right.\left|\Psi\right\rangle \\
    &= i\overset{\infty}{\underset{0}{\int}}dE\frac{d}{d\varepsilon}\left.\left(\left\langle t,b\right.\left|E+\varepsilon,f\right\rangle \right)\right|_{\varepsilon=0}\left\langle E,b\right.\left|\Psi\right\rangle \\
    &= t\overset{\infty}{\underset{0}{\int}}dE\frac{e^{-iEt}}{\sqrt{2\pi}}\left\langle E,b\right.\left|\Psi\right\rangle =t\left|\Psi^{b}\left(t\right)\right\rangle 
\end{split}    
\end{equation}

Thus, the action of the fixed-point time operator defined in Eq. \eqref{eq:t_FP} is given by:

\begin{equation}
\begin{split}
    \hat{t}^{FP}\left\llbracket \Psi\right\rrbracket _{t} &= \frac{1}{2} \left[\hat{t}^{b}\left|\Psi^{b}\left(t\right)\right\rangle \otimes\left|\Psi^{f}\left(t\right)\right\rangle +\left|\Psi^{b}\left(t\right)\right\rangle \otimes\hat{t}^{f}\left|\Psi^{f}\left(t\right)\right\rangle \right] \\
    &= \frac{1}{2} \left[\left\langle t\right|\hat{t}\left|\bar{P}\Psi\right\rangle \otimes\left|\Psi^{f}\left(t\right)\right\rangle +\left|\Psi^{b}\left(t\right)\right\rangle \otimes\left\langle t\right|\hat{t}\left|P\Psi\right\rangle \right] \\
    &= t\left\llbracket \Psi\right\rrbracket _{t}
\end{split}
\end{equation}

\end{widetext}

%\subsection{Observables, expectation values and weak values}
%This subsection builds on the previous one in evolving the initial/final states forward/backward in time, but we still have to understand how an arbitrary observable $A$ looks like within the various formulations (mainly in Bauer's). Is it doubled exactly like the Hamiltonian?  

\subsection{Conceptual Connections}\label{sec:Conceptual}

In this section we expand on the conceptual foundations and consequences of each theory considered above, to illustrate the choices one is faced with when deciding between two times or none. A full taxonomy of temporal properties discussed here is shown in Table \ref{table:TTON}.

At first sight, the PW formalism seems to trivially treat all times equivalently. The WDW equation dictates a universe with frozen dynamics. There is no explicit time-dependency and therefore we may argue that at the birds-eye level of the total system this description is trivially both event-symmetric and time-symmetric. However, once we divide the total system into a clock subsystem and the rest, the dynamics of the latter could be time-asymmetric (see e.g. the case of effective non-Hermitian dynamics \cite{paiva2022non,cohen2023quantum}) and event-asymmetric from the reduced perspective of the clock subsystem, which follows the `one-way' dynamics shown in Fig. \ref{fig:one-way}. Since within this framework, times are defined in terms of the clock states, our classification of the temporal properties of the PW approach shown in Table \ref{table:TTON} reflects the reduced perspective only.

The history states in the expansion of Eq. \eqref{eq:PW_History} are often interpreted as time-extended, mutually exclusive worlds, in the Everettian sense \cite{favalli2022peaceful,kuypers2022quantum,kuypers2024measuring,favalli2024emergence}. The many worlds interpretation (MWI) of Everett \cite{everett_relative_1957} is therefore most often the interpretational framework in which to the PW approach is understood \cite{deutsch_apart_2010}. The construction of a universal wavefunction containing many (real or potential) quantum histories seems to rule out collapse-based interpretations from the outset, and to favour the deterministic, unitary-only prescription of the MWI. However, Adlam argues against this orthodoxy in Ref. \cite{adlam2022watching}, arguing that, since the MWI cannot yet agree on how to account for probabilities, an Everettian interpretation of the PW approach, which must at some point assign probability measures over quantum histories, forces the PW approach to suffer from the same problems. Adlam advocates for a single-world, realist account of the PW and IQRF programmes, favouring Kent's formulation of a `final-measurement' solution \cite{kent2014solution,kent2015lorentzian,kent2017quantum}. In this interpretation a post-selection measurement on the universal wavefunction, deferred in a limiting process to the infinite future, determines the records in the particular history state which is actualized. However, this theory requires that, in the infinite future, all particles and fields will have ceased interacting - a very strong assumption indeed. Additionally, although it might arguably be possible to make sense of probability assignments in this interpretation, the `final-measurement' proposal gets us no closer to actually assigning a mathematical form to those probabilities. Finally, it is difficult to justify the occurrence of a final measurement without also committing to either a fundamentally stochastic evolution or to an external system which acts upon the universal wavefunction to carry out this measurement.

Another type of `one-way' histories framework is the decoherent histories formulation; popular amongst proponents of the MWI, who view it as a formalization of the idea of multiple wavefunction branches evolving in parallel \cite{deutsch_quantum_1985,saunders_branching_2008,wallace_emergent_2012,strasberg2024first,wang2025decoherence}. On this account, the existence of a fixed boundary condition at the initial time $t_{1}$ means that the universal wavefunction has temporal asymmetry built into it - it undergoes branching in the direction of increasing entropy \cite{gell-mann_quantum_1996,riedel_classical_2017}, as illustrated schematically in Fig. \ref{fig:CH_Psi}. The histories formulation also offers a resolution of the problem of identity of an object over time - even if some temporal parts overlap in a pair of quantum histories describing that object, the different trajectories in time can be mutually exclusive if they contain orthogonal parts at one time in the sequence \cite{wallace_emergent_2012}. This is also shown in Fig. \ref{fig:CH_Psi}, where distinct histories may coincide from time $t_{1}$ to $t_{2}$, at which point they diverge. If an observer is identified with the entire history, they can legitimately be uncertain about which entire history they will discover themselves to be in when they conduct a measurement after the branching has occurred, and can meaningfully assign probabilities to being located within one entire history or another \cite{wilson_everettian_2012}. This arguably resolves the problem of uncertainty about the future (also known as the incoherence problem) within the deterministic unitary evolution of the MWI \cite{saunders_branching_2008,wilson_objective_2013}, but does not on its own resolve the quantitative problem of explaining the Born measure.

In all of the `two-time' quantum formalisms or interpretations discussed above, much depends on whether the reverse direction of time is taken to be an intrinsic feature of the physical world, or a mere mathematical convenience. A closely related question is whether, in time-symmetric theories, the reverse arrow of time carries physical information and can be described as `causal' in any realistic sense. However, recent strong results in Refs. \cite{price_does_2012,leifer2017time} indicate that time symmetry indeed implies retrocausality, in the sense of a physical influence propagating backwards in time.

The Bauer approach uses the pseudospin (or equivalently, negative energy) extension of standard quantum theory without further interpretation - Bauer's goal is to construct a time operator without committing to an ontic interpretation of the extended part of the state space. Indeed, Bauer refers to the construction of the `physical space' as the space obtained by projecting out the degrees of freedom corresponding to backwards time solutions \cite{bauer1983time}.

Whereas the Bauer approach is manifestly time- and event-symmetric, the TSVF maintains time symmetry but violates event symmetry \cite{ridley2025time}. This is because, taken as a representational model of reality, it divides time into past, present and future at the level of the mathematical structures used to describe these regions of time. Specifically, in the case of a measurement at time $t$ with pre- and post-selection at times $t_{1}$ and $t_{2}$, the future-directed component of the two-state vector exists in the past of $t$, with no backwards propagation in the region $\left[t_{1},t\right]$, and the backwards-directed component of of the two-state vector exists in the future of $t$, with no forwards propagation in the region $\left[t,t_{2}\right]$. This type of propagation corresponds to the `corridor' dynamics shown in Fig. \ref{fig:corridor}. In addition, the TSVF allows overlap to occur between the forwards- and backwards-propagating parts of the two-state, which theoretically exist in the standard Hilbert space $\mathcal{H}_{t_{1}}$ and the dual spce $\mathcal{H}^{\dagger}_{t_{2}}$, respectively, before being mapped via unitary evolution to the pair of Hilbert spaces $\mathcal{H}_{t}$, $\mathcal{H}^{\dagger}_{t}$.

According to Vaidman in Ref. \cite{vaidman_time_2010}, the two directions of propagation in the TSVF also come with an interpretational distinction. In this view, the forward-propagating part of the two-state vector is interpreted ontologically in accordance with the MWI, whereas the backwards-propagating part of the two-state vector is given an epistemic interpretation. Vaidman's position was defended in Ref. \cite{robertson2017can}, where it is argued that in practice there does seem to be a difference in function between the forwards- and backwards-directed components; in particular, the backwards-directed parts do not have the same `causal power' as the forwards-directed parts in determining a single measurement outcome at the measurement time $t$. For this, the forwards-directed component, taken in isolation, is arguably both necessary and sufficient.

However, in addition to the TSVF and its multiple-time generalization, an interpretation of quantum theory with fundamental pre- and post-selection has arisen in recent years, the so-called \emph{two-time interpretation} (TTI)
\cite{aharonov2005two,aharonov2014measurement}. According to the TTI, there really exists a backward travelling wavefunction from the future, which overlaps with the pre-selected state at the strong measurement time $t$, supplying a mechanism for the collapse of the wavefunction \cite{aharonov2014measurement}. This is carried out via a selection of a special `destiny' vector serving as a final boundary condition on the universe, which is chosen such that it guarantees Born rule statistics at the measurement time \cite{aharonov_two-time_2017}. The TTI therefore provides a reasonable ontic interpretation of the entire two-state vector whilst violating event symmetry \cite{ridley2025time}. It is also not clear, in this formulation, how to construct a time operator - it does not map onto Bauer's approach because the direction of propagation of the two-state vector flips sign at the measurement time. 

Moving on to ETNU, the multiple-time extension of the TSVF discussed in Section \ref{sec:TSVF}, we note that, since the ETNU formalism breaks up time into a sequence of fundamentally BTFP propagations, it violates event-symmetry, whilst maintaining time symmetry in its basic dynamics. As such, the temporal properties used to classify the TSVF and ETNU approaches are identical as shown in Table \ref{table:TTON}.

One of the questions motivating ETNU is the validity of an ergodic-like property of quantum mechanics. In particular, Ref. \cite{aharonov_each_2014} begins by attempting to build up an $N_{t}$-time state in the same way as one would construct the many-body state composed of $N_{t}$ particles. This is shown not to be possible, as the resulting multipartite-like state cannot account for all the correlations seen in a genuine multiple-time quantum measurement. Essentially, this is because each time may correspond to \emph{either} a forward-propagating pre-selection \emph{or} a backwards-time-directed post-selection. 

The status of the time domain in the TI is a somewhat nuanced question. The TI describes measurements in terms of a two-time `handshake' process that occurs across the entire temporal region separating the absorber from the emitter, such that a measurement is not confined to a single point in time but is rather a fundamentally time nonlocal process. As such, the TI could be argued to involve the `motorway' two-time dynamics shown in Fig. \ref{fig:motorway}. However, Cramer claimed that the individual offer and confirmation waves occurred in what he terms `pseudotime' and are not physical processes with separate arrows of time \cite{cramer_transactional_1986}. This makes it difficult to see how a time operator may be constructed in the TI along the lines of Bauer. On Cramer's account, the only real processes which occur in nature are completed transactions, which can be conceived of as four-dimensional standing waves occurring along the regular timeline shown in Fig. \ref{fig:one-way}, although it is still in some sense a `motorway' theory. Maudlin has criticized the pseudotime account as incoherent \cite{maudlin2011quantum}. He also describes an experiment in which the absorbers of the TI have a contingent location - in this case the TI seems to give paradoxical results \cite{maudlin2011quantum,lewis2013retrocausal,ridley2025time}. In addition, the TI as originally formulated by Cramer is not event-symmetric - it involves privileged points in time corresponding to the location of absorbers and emitters. This could only be overcome by placing absorber/emitter structures at \emph{every} point in time, which makes it exceedingly difficult to account for single measurement outcomes \cite{lewis2013retrocausal,ridley2025time}.

To meet the criticisms of Maudlin, an alternative account of the TI has been developed by Kastner - the so-called \emph{possibilist} transactional interpretation, or PTI \cite{kastner2006cramer,kastner2012possibilist,kastner_transactional_2013,kastner2016transactional,kastner2017there}. Kastner interprets the TI within a modal realist ontology. According to Kastner, the term `reality' is not confined to processes occurring in spacetime. The PTI states that the quantum processes corresponding to the offer and confirmation waves occur in a `pre-spacetime realm of possibility' \cite{kastner2012possibilist}, which impinges upon the empirically observable spacetime realm when a completed transaction occurs. This meets the first of Maudlin's criticisms, however Lewis has argued compellingly that it does not meet the second criticism, unless we revert to a `baroque' version of the MWI, where all possible transactions occur on different branches of the universal wavefunction \cite{lewis2013retrocausal}. 

The FPF offers a novel vision of reality, as fundamentally composed of two-time channels connecting states which exist at different times, and which are connected along both time directions. It therefore falls squarely into the `motorway' category of time propagation shown in Fig. \ref{fig:motorway}, and is both time- and event-symmetric \cite{ridley2025time}. The FPF also offers a picture of measurement - measurements are physical procedures which serve to \emph{locate the observer} within the seemingly immutable object called the universal wavefunction $\left|\Psi_{U}\right\rangle$ \cite{ridley2023quantum}. When a sequence of $N_{t}$ strong measurements is carried out, the full universal wave function is effectively constrained such that the region of reality accessible to the observer is a superposition of histories consistent with the outcomes of those measurements. The act of measurement fixes the state on both branches of the Keldysh contour and is therefore represented by a fixed point, which is mathematically defined in Eq. \eqref{eq:1FP} and illustrated in Fig. \ref{fig:1FP}. This is a state of maximum order and corresponds to a crossing point for quantum histories. As such, it provides a quantum counterpart to the relativistic notion of an event as a point where worldlines cross. In this theory, the ontology is not composed of events - the wavefunction is ontologically primitive. However, a notion of an event which fulfils several of Maccone's criteria in Ref. \cite{maccone_fundamental_2019} for a quantum theory of events is constructed from the usual quantum language of state vectors. 

Crucially, within the FPF a concept of probability is given, which has been extensively defended and connected to longstanding debates in the philosophy of probability in Ref. \cite{ridley2025MRW}. From this concept of probability, an identification of quantum probability with the measure of existence of a quantum history is made. Then it can be shown mathematically that the measure of existence is equal to the Born rule \cite{ridley2023quantum,ridley2025MRW}. This offers a derivation of the Born rule from the minimal number of assumptions. Recently, the FPF has been given an interpretational framework - many retrocausal worlds (MRW) - which can be construed as a fully time-symmetric version of the MWI, with branching in both time directions, and `worlds' understood as time-extended world-tubes within a diverging picture of the branching process \cite{ridley2025MRW}. An example of a constrained `universe' in the MRW view is shown in Fig. \ref{fig:ABL_Compact}. 

Note that the fixed point idea is distinct from the Janus point concept used by Barbour \cite{barbour2014identification,barbour2020janus} to describe forwards/backwards time propagation after/prior to the initial condition of the universe - in Barbour's theory, there is \emph{only} FPTB propagation. This type of propagation is not shown on Fig. \ref{fig:schematic}, but can be visualised as an `anti-corridor' configuration. In the FPF, there is no single arrow of time in any temporal region, but two time orientations at every moment. As such, every moment experienced by the observer is fundamentally composed out of future and past-directed parts, and each branch in the past of a fixed point is likewise composed of temporal parts on both branches of the Keldysh contour shown in Fig. \ref{fig:Time_contour}.

\begin{widetext}

\begin{table}[htbp]
\begin{tabular}{|p{2cm}|| >{\centering\arraybackslash}m{2cm} | >{\centering\arraybackslash}m{2cm} | >{\centering\arraybackslash}m{2cm} | >{\centering\arraybackslash}m{2cm} | >{\centering\arraybackslash}m{2cm} | >{\centering\arraybackslash}m{2cm} |}
 \hline
 \multicolumn{7}{|c|}{Temporal Properties} \\
 \hline
  & Event Symmetry & Time Symmetry & One-way & Corridor & Motorway & Time Operator \\
 \hline
 \hline
 PW   & \xmark    & \xmark & \cmark & \xmark & \xmark & \cmark\\
 \hline
 Bauer &   \cmark  & \cmark  & \xmark & \xmark & \cmark & \cmark \\
 \hline
 TSVF & \xmark & \cmark & \xmark & \cmark & \xmark & \xmark \\
 \hline
 ETNU    & \xmark & \cmark &  \xmark & \cmark & \xmark & \xmark \\
 \hline
 TI &   \xmark  & \cmark & \xmark & \xmark & \cmark & \cmark \\
 \hline
 FPF & \cmark  & \cmark & \xmark & \xmark & \cmark & \cmark \\
 \hline
\end{tabular}

\caption{Comparison of the different formalisms considered in virtue of their basic temporal properties.}
\label{table:TTON}
\end{table}

\end{widetext}

We summarize our discussion of the fundamental temporal properties satisfied by the various formalisms considered in this work in Table \ref{table:TTON}; there, we list the symmetries and type of time propagation that hold in each theory. We also show which theories admit the construction of a time operator.

\subsection{Two Times AND None?}

We note that within the Bauer approach, decoherent histories, TSVF, MTS, ETNU and FPF, time is introduced as an external parameter (parameterizing either the state or the Hilbert space), just like within standard quantum mechanics. Evolution of wavefunctions in both time-directions is possible, and moreover has a deep meaning in these approaches, but even though a time operator can be constructed in some of these approaches, time itself is not introduced as a dynamical variable like time in general relativity. Moreover, these approaches all appeal to a notion of absolute time, whereas the PW approach quantizes the constraint equation and projects the universal state onto a subsystem in order to recover \emph{relational} dynamics. Therefore, it is tempting to time-symmetrize the latter framework or alternatively, promote time in the former frameworks to an operator. In many ways, this is similar to recent analyses, e.g. \cite{paiva2022flow,paiva2022non}, where more than one quantum clock was used within the PW formalism, leading to different effective dynamics as seen from the perspective of each clock through effective equations of motion. In this sense, the PW approach, and its generalization towards spatio-temporal reference frames \cite{suleymanov2024nonrelativistic}, go much beyond two time approaches -- they readily support the inclusion of multiple clocks, fundamentally adhering to a {\it relational} description of quantum mechanics \cite{hohn2021trinity,adlam2022watching}, quantum field theory \cite{hohn2021equivalence,oliveira2025spacetime} and group field theory \cite{calcinari2025relational}. To enjoy these benefits of the PW formalism alongside with the inherent time-symmetry of two time approaches (potentially including their treatment of nonlocality \cite{sutherland2008causally,sutherland1983bell,price2015disentangling,wharton2024localized,wharton2020colloquium,hall_measurement-dependence_2020}) we propose a united formulation which treats the two times in an operational manner as two PW clocks with opposite directions \cite{ridley2025twospin}.

This will involve the introduction of \emph{two} WDW constraint equations for the universal state vector

\begin{gather}
    H^{f}_{T}|\Psi\rangle\rangle = 0 \\ 
    H^{b}_{T}|\Psi\rangle\rangle = 0,
\end{gather}

where the `$f$' and `$b$' labels denote the upper and lower branches of the Keldysh contour, on which an emergent reduced two-time dynamics can be defined.

\section{Conclusions}

Traditionally, time in quantum mechanics has been introduced according to the thesis that it is a single background parameter. Modifications of this thesis have mostly focused on making time an emergent phenomenon, i.e. they have replaced the Newtonian concept of absolute time with a notion of time that emerges from the entanglement between subsystems in a closed universe. However, there is a second clause to this thesis, namely that there is only a single time parameter, which has not received the same level of attention as the first. We have argued that, with a view to resolving at least one aspect of the problem of time - the construction of a time operator - it can be just as advantageous to drop the second clause of this assumption as the first clause. In other words, there appears to be a choice that one can make between dropping the `background' aspect of quantum time, and dropping the `singular' aspect. 

We have thus examined approaches which do the latter, through the introduction of two or more temporal degrees of freedom. In particular, we have shown how a second temporal degree of freedom resolves conceptual and formal problems that appear when treating time in the traditional way. Some of these problems, such as the existence of a time operator, are tightly connected to the traditional problem of time. Other advantages of the two-time perspective, such as the possibility of deriving the Born rule, are not usually associated with the problem of time, but are no less important when considering whether or not to move to this perspective.

At first, we have presented a framework and a set of criteria for making a \emph{choice} between the PW and various two-time approaches, i.e. a choice between dropping the assumption that time exists in the background and the assumption that time is singular.

However, a third option is available, namely that we drop \emph{both} assumptions and move to a picture of time which is both emergent and amenable to the two-time treatment. This motivates our construction, in Ref. \cite{ridley2025twospin}, of a PW spin clock with two emergent temporal degrees of freedom, possessing opposite orientations. This is not intended as a model of reality itself, but as a model which is isomorphic to the Keldysh structure one should see in nature if time is indeed both emergent and non-singular. 

Although this work focuses on fundamental questions related to time in quantum mechanics, we find it plausible that the formulation developed here could enhance quantum-limited timekeeping and networked clock metrology \cite{Ludlow2015RMP,Pezzè2018RMP}, as well as clock synchronization \cite{Jozsa2000PRL,Giovannetti2001Nature,Komar2014NatPhys}, with further implications for quantum computation and quantum communication via superpositions of temporal order \cite{Oreshkov2012NatComm,Ebler2018PRL,Chiribella2022CommsPhys}.

Finally, we note that the approaches discussed in this paper have all been non-relativistic. However, to address the problem of time in quantum gravity it will be necessary to extend them to relativistic settings \cite{diaz2019history,hohn2021equivalence, oliveira2025spacetime}. 

\textit{Acknowledgments}.--
The authors wish to thank Nathan Argaman, Ismael L. Paiva, Michael Suleymanov and Ken Wharton for helpful comments and discussions. The authors were supported by the European Union’s Horizon Europe research and innovation programme under grant agreement No. 101178170 and by the Israel Science Foundation under grant agreement No. 2208/24. This research was supported in part by Perimeter Institute for Theoretical Physics. Research at Perimeter Institute is supported by the Government of Canada through the Department of Innovation, Science and Economic Development and by the Province of Ontario through the Ministry of Research, Innovation and Science.

\bibliographystyle{apsrev4-1}

\end{document}